\documentclass[reprint,showpacs,preprintnumbers,pre,superscriptaddress]{revtex4-1}
\usepackage[T1]{fontenc}
\usepackage{textcomp}
\usepackage[latin9]{inputenc}
\usepackage{array}
\usepackage{amsmath}
\usepackage{amssymb}
\usepackage{graphicx}

\makeatletter

\providecommand{\tabularnewline}{\\}

\usepackage{bm}\usepackage{amsfonts}
\usepackage{color}%\usepackage{stfloats}

\makeatother

\begin{document}
\title{Shortcuts to state transitions for active matter}
\author{Guodong Cheng}
\affiliation{School of Physics and Astronomy, Beijing Normal University, Beijing
100875, China}
\author{Z. C. Tu}
\affiliation{School of Physics and Astronomy, Beijing Normal University, Beijing
100875, China}
\author{Geng Li}
\email{gengli@bnu.edu.cn}

\affiliation{School of Systems Science, Beijing Normal University, Beijing 100875,
China}
\begin{abstract}
Shortcut schemes can accelerate quasi-static processes in passive
systems by adding auxiliary controls to realize swift transitions
between equilibrium states. In active systems, however, inherently
directed motion driven by free energy consumption continually drives
the system away from equilibrium. In this work, we develop a shortcut
framework to realize swift state transitions for active systems operating
in the weak activity regime. An auxiliary potential is introduced
to guide the system along a predefined distribution path, allowing
it to reach the target state within a finite time. Considering unavoidable
energy cost in such a finite-time process, we derive a thermodynamic
metric from the dissipative work to induce a Riemann manifold on the
space spanned by the control parameters. The optimal protocol with
minimum dissipative work is then identical to the geodesic path in
the geometric space. We demonstrate this framework by considering
active systems confined in an external harmonic trap and interacting
via two distinct internal potentials, respectively: an attractive
harmonic coupling and a repulsive pairwise Gaussian-core coupling.
The strengths of both the external trap and the internal interactions
are controllable. For the latter case, since the auxiliary potential
can not be derived precisely, we adopt a variational method to obtain
an approximate auxiliary control. Compared to linear protocols, the
geodesic protocols can effectively reduce dissipation.
\end{abstract}
\maketitle

\section{INTRODUCTION}

Active matter constitutes a class of non-equilibrium systems that
consume free energy to generate directed motion. These systems are
abundant in nature, exemplified by bacterial suspensions, bird flocks,
and fish schools \cite{bacterial,bird_flock,flocking_1,flock_2,MIPS}.
Inspired by these living systems, researchers have engineered diverse
artificial active matter, including Janus particles \cite{janus_particle},
active colloids \cite{active_colloids} and microrobot swarms \cite{microrobot_swarms}.
These synthetic systems offer distinct advantages over their biological
counterparts, particularly in terms of their controllability and flexibility
\cite{ABP_2}. They can serve as experimental platforms for investigating
non-equilibrium statistical physics \cite{ABP_2,experimental_platform},
and act as carriers for realizing in-vivo imaging \cite{vivo_images},
intelligent materials \cite{smart_material_1}, efficient active engines
\cite{active_engines_1,active_engines_2}, and environmental technologies
like water purification \cite{water_cleaning}. Across all these applications,
achieving swift and efficient state transitions remains a critical
yet challenging objective.

The problem of realizing swift state transitions has been thoroughly
investigated in passive systems. One of the mainstream solutions is
to introduce an additional flow field or potential to influence the
dynamics of the system \cite{vaikuntanathan2008escorted,vaikuntanathan2011escorted,speed_up,speeding_up_relaxation_time,shortcut_Li,shortcut_1,shortcut_2}.
The modified dynamics can generate evolutionary paths or reach target
states which are designed for various tasks that would otherwise take
a longer, or even infinite, time under the original dynamics. The
strategy has been used in generating near-equilibrium paths for efficient
and fast free energy estimates \cite{vaikuntanathan2008escorted,vaikuntanathan2011escorted},
speeding-up the relaxation time \cite{speeding_up_relaxation_time,speed_up},
realizing isothermal process beyond a quasi-static process \cite{shortcut_Li}.
However, extending this approach to active matter raises the question
of how to overcome the distinct challenges posed by its inherent non-equilibrium
character and complex interparticle interactions.

A pivotal consideration in a swift state-transition process is the
unavoidable energy cost, which can be minimized by various schemes.
Among the most systematic approaches is thermodynamic geometry \cite{MeasuringThermodynamicLength,Thermodynamic_Metrics_and_OptimalPaths,Geometry_Li,wang2024thermodynamic,p2gz-47vt}.
In this framework, dissipative work is expressed geometrically through
a positive semi-definite metric tensor, known as the thermodynamic
metric. Consequently, the optimal protocol yielding minimum dissipative
work corresponds to the geodesic path on the Riemannian manifold of
control parameters defined by this metric. Alternatively, optimal
transport theory employs the Wasserstein distance to quantify the
transition cost between initial and target states to determine the
minimal-cost path \cite{opt_1,opt_2}. These two frameworks become
equivalent provided the control parameters are sufficiently expressive
\cite{Geometry_vs_optimalTransport,ito2024geometric}.
\begin{figure}
\centering{}\includegraphics[width=0.8\columnwidth,keepaspectratio,height=0.8\columnwidth]{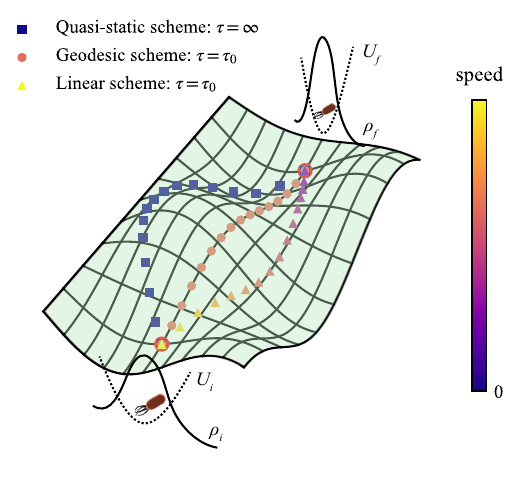}\caption{\label{fig:framework}Schematic illustration of the shortcut framework
for active systems. The curved surface represents the Riemannian manifold
induced by the thermodynamic metric in the space of control parameters.
Adjacent to the start point (initial state) and end point (final state)
are small diagrams showing the potential energy curve $U$ (dashed
line) and the probability distribution $\rho$ (solid line), with
an E. coli bacterium at the potential minimum. The diagrams are labeled
with $U_{i}$, $\rho_{i}$ and $U_{f}$, $\rho_{f}$, respectively.
Three representative transition paths connect the start and end points,
with colors encoding the instantaneous speed along the path as indicated
by the color bar. The curve with zero speed (as indicated by the color
bar) corresponds to the quasi-static scheme ($\tau=\infty$), where
the process is infinitely slow. The curve with a uniform nonzero color
represents the geodesic scheme ($\tau=\tau_{0}$), which follows the
optimal path with constant speed in the metric space. The third curve,
with varying colors, illustrates a linear driving scheme ($\tau=\tau_{0}$)
as a reference non-optimal protocol.}
\end{figure}

In this work, we develop a shortcut approach to realize swift state
transitions for active systems--specifically, systems of interacting
active Ornstein-Uhlenbeck particles \cite{AOUPs}--in the weak activity
regime. As shown in Fig. \ref{fig:framework}, energy cost of such
a process can be optimized by using thermodynamic geometric scheme.
In Sec. \ref{sec:Shortcuts}, an auxiliary potential is introduced
to steer the system along a predefined evolutionary distribution connecting
the initial and the final distributions. The derivation of this potential
relies on inversely solving the evolution equation of probability
distribution for active matter, which can be derived by applying the
Fox approximation \cite{FOX_multidimensional_diagnal,FOX,FOX_multidimensional_offdiagnal}.
Since for complex interactions the precise auxiliary potential cannot
be derived analytically and traditional numerical methods face exponential
cost due to the curse of dimensionality, in Sec. \ref{sec:Approximate-shortcut},
we develop a variational approach to obtain an approximate auxiliary
control. Specifically, we define a series of functionals whose stationary
conditions are equivalent to the evolution equation governing the
auxiliary control. Then some parameterized ansatzes are introduced
to represent the auxiliary control so that the stationary conditions
yield a system of algebraic equations in which coefficients can be
evaluated by sampling techniques. Finally, solve the algebraic equations
to obtain an approximate auxiliary control. In Sec. \ref{sec:energy-cost},
to minimize the energy cost in the process, we express the dissipative
work in a geometric form in which a positive semi-definite metric
tensor is defined, referred to as the thermodynamic metric. The optimal
protocol with minimum dissipative work corresponds to the geodesic
path in the Riemannian manifold of control parameters, equipped with
the thermodynamic metric. In Secs. \ref{subsec:harmonic_coupling}
and \ref{subsec:gaussian_coupling}, we consider two active systems
to show the flexibility of our method across different interaction
types. The first system involves particles coupled through an attractive
harmonic potential, whereas the second involves particles interacting
via a repulsive pairwise Gaussian-core potential. This complementary
choice effectively highlights the adaptability of our framework.

\section{Theoretical model for active matter}

We study an active system of $N$ self-propelled particles subject
to mutual interactions and an external potential. The total potential
$U(\boldsymbol{r},t)$ comprises these two contributions. The microstate
of the system is described by $\boldsymbol{r}\equiv\left(\boldsymbol{r}_{1},\boldsymbol{r}_{2},\dots,\boldsymbol{r}_{N}\right)^{\text{T}}$,
where $\boldsymbol{r}_{i}$ represents the position of particle $i$.
There exist various models to describe the dynamics of active systems;
notably Active Brownian particles(ABPs) \cite{ABP_1,ABP_2} and Active
Ornstein-Uhlenbeck particles (AOUPs) \cite{AOUPs}. Here we adopt
the AOUP model, as it is more amenable to analytical treatment, particularly
in the presence of interactions. The dynamics of particle $i$ subject
to Stokes friction with coefficient $\zeta$ can be described by an
overdamped Langevin equation

\begin{equation}
\zeta\dot{\boldsymbol{r}}_{i}(t)=-\nabla_{i}U+\boldsymbol{\xi}_{i}(t)+\boldsymbol{\eta}_{i}(t),\label{eq:AOUP}
\end{equation}
where $\nabla_{i}$ denotes the gradient with respect to the position
$\boldsymbol{r}_{i}$ of particle $i$. We have considered both the
Gaussian white noise and Gaussian colored noise, both with zero mean
and variances: $\langle\xi_{i}^{\alpha}(t)\xi_{j}^{\beta}(t')\rangle=2\zeta\delta^{\alpha\beta}\delta_{ij}\delta(t-t')$
and $\langle\eta_{i}^{\alpha}(t)\eta_{j}^{\beta}(t')\rangle=(\gamma D_{a}/\tau_{p})\delta^{\alpha\beta}\delta_{ij}\mathrm{e}^{-|t-t'|/\tau_{p}}$,
where the Greek superscripts $\alpha$, $\beta$ denote vector components.
The noise $\boldsymbol{\xi}$ represents thermal fluctuations arising
from the solvent, while the exponentially correlated noise $\boldsymbol{\eta}$
encodes persistence and thus captures the nonequilibrium nature of
activity. Its statistics are controlled by the noise amplitude $D_{a}$
and the persistence time $\tau_{p}$, which are typically proportional,
$D_{a}\propto\tau_{p}$. Throughout this work, we set $\zeta$ and
$k_{B}T$ to unity.

The non-Markovian character of AOUPs prevents a direct derivation
of a time evolution equation for the system probability distribution.
To overcome this, one can employ approximate schemes that map the
dynamics onto an effective Markovian description, such as the unified
colored noise approximation (UCNA) \cite{maggi2015multidimensional}
and the Fox approximation \cite{FOX,FOX_multidimensional_diagnal,FOX_multidimensional_offdiagnal}.
We employ the latter method due to its superior accuracy when translational
white noise $\boldsymbol{\xi}$ is present \cite{FOX_vs_UCNA}. And
the time evolution of probability distribution $\rho(\boldsymbol{r},t)$
is governed by the following equation:

\begin{equation}
\frac{\partial\rho(\boldsymbol{r},t)}{\partial t}=-\nabla_{i}\cdot\boldsymbol{J}_{i}(\boldsymbol{r},t),\label{eq:FP}
\end{equation}
where the probability current \textbf{$\boldsymbol{J}$} is given
by

\begin{eqnarray}
J_{i}^{\alpha} & = & -\left(\nabla_{i}^{\alpha}U\right)\rho(\boldsymbol{r},t)-\nabla_{i}^{\alpha}\rho(\boldsymbol{r},t)\nonumber \\
 &  & -D_{a}\nabla_{j}^{\beta}\left[(\mathbf{I}+\tau_{p}\nabla\nabla U)^{-1}\right]_{ij}^{\alpha\beta}\rho(\boldsymbol{r},t).\label{eq:current_expression}
\end{eqnarray}
Throughout this work we employ the Einstein summation convention over
repeated indices. A complete derivation of this equation is provided
in Supplementary Material \cite{sm}. Here, we retain the off\nobreakdash-diagonal
components in the last term of Eq. (\ref{eq:current_expression})
which represent many\nobreakdash-body couplings, thereby preserving
collective interactions more faithfully \cite{FOX_multidimensional_diagnal,FOX_vs_UCNA}.
For short persistence times, the current can be expanded perturbatively
in $\tau_{p}$. To first order, one obtains $J_{i}^{\alpha}=-\left\{ \left(\nabla_{i}^{\alpha}U\right)\rho+(1+D_{a})\nabla_{i}^{\alpha}\rho\right\} $,
which will be referred to hereafter as the first\nobreakdash-order
approximation. Extending the expansion to second order yields $J_{i}^{\alpha}=-\left\{ \left(\nabla_{i}^{\alpha}U\right)\rho+(1+D_{a})\nabla_{i}^{\alpha}\rho\right\} -D_{a}\tau_{p}\nabla_{j}^{\beta}\left(\nabla_{i}^{\alpha}\nabla_{j}^{\beta}U\right)\rho$,
and will be termed the second\nobreakdash-order approximation.

\section{Shortcuts scheme for realizing swift state transitions\label{sec:Shortcuts}}

We aim to drive the system from an initial distribution $\rho_{i}$
to a target distribution $\rho_{f}$. To this end, we prescribe an
interpolation path $\rho_{B}(\boldsymbol{r},\boldsymbol{\lambda}(t))\equiv\exp\{\frac{F(\boldsymbol{\lambda}(t))-U_{o}(\boldsymbol{r},\boldsymbol{\lambda}(t))}{1+D_{a}}\}$,
with boundary conditions $\rho_{i}=\rho_{B}(\boldsymbol{r},\boldsymbol{\lambda}(0))$
and $\rho_{f}=\rho_{B}(\boldsymbol{r},\boldsymbol{\lambda}(\tau))$.
Here, $U_{o}(\boldsymbol{r},\boldsymbol{\lambda}(t))\equiv-(1+D_{a})\log\rho_{B}(\boldsymbol{r},\boldsymbol{\lambda}(t))$
defines the original potential. The vector $\boldsymbol{\lambda}(t)$
represents a set of tunable parameters, $\tau$ denotes the total
driving time, and $F$ is a normalization factor. To guide the system
evolution along the predefined path, an auxiliary potential $U_{a}$
must be introduced.

\subsection{First-order approximation}

Under the first-order approximation, substituting $\rho(\boldsymbol{r},t)=\rho_{B}(\boldsymbol{r},\boldsymbol{\lambda}(t))$
into Eq. (\ref{eq:FP}) yields:
\begin{equation}
\frac{\partial\rho_{B}}{\partial t}=\nabla_{i}\cdot\left\{ \left(\nabla_{i}(U_{o}+U_{a}^{(1)})\right)\rho_{B}+(1+D_{a})\nabla_{i}\rho_{B}\right\} ,
\end{equation}
where $U_{a}^{(1)}$ denotes the auxiliary potential under the first-order
approximation. After simplification, we obtain the governing partial
differential equation:
\begin{equation}
(1+D_{a})\nabla^{2}U_{a}^{(1)}-\nabla_{i}U_{a}^{(1)}\cdot\nabla_{i}U_{o}=\left(\frac{\partial F}{\partial\lambda_{\mu}}-\frac{\partial U_{o}}{\partial\lambda_{\mu}}\right)\dot{\lambda}_{\mu}.\label{eq:U_a_1_pde}
\end{equation}
From its structure, the auxiliary potential admits the general form
\begin{equation}
U_{a}^{(1)}(\boldsymbol{r},\boldsymbol{\lambda},\dot{\boldsymbol{\lambda}})=\dot{\boldsymbol{\lambda}}\cdot\boldsymbol{f}(\boldsymbol{r},\boldsymbol{\lambda}).\label{eq:U_a_1_form}
\end{equation}
Substituting this ansatz yields the partial differential equation
\begin{equation}
D_{\mu}^{(1)}(f_{\mu})\equiv(1+D_{a})\nabla^{2}f_{\mu}-\nabla_{i}f_{\mu}\cdot\nabla_{i}U_{o}+\frac{\partial U_{o}}{\partial\lambda_{\mu}}-\frac{\partial F}{\partial\lambda_{\mu}}=0.\label{eq:f_pde}
\end{equation}

\subsection{Second-order approximation}

Proceeding to the second-order approximation and substituting $\rho(\boldsymbol{r},t)=\rho_{B}(\boldsymbol{r},\boldsymbol{\lambda}(t))$
into Eq. (\ref{eq:FP}), we obtain 
\begin{eqnarray}
\dot{\lambda}_{\mu}\frac{\partial\rho_{B}}{\partial\lambda_{\mu}} & = & -D_{a}\tau_{p}\frac{\partial^{2}}{\partial r_{i}^{\alpha}\partial r_{j}^{\beta}}(\frac{\partial^{2}(U_{o}+U_{a}^{(2)})}{\partial r_{i}^{\alpha}\partial r_{j}^{\beta}}\rho_{B})\nonumber \\
 &  & +\frac{\partial}{\partial r_{i}^{\alpha}}\left(\frac{\partial(U_{o}+U_{a}^{(2)})}{\partial r_{i}^{\alpha}}\rho_{B}\right)+(1+D_{a})\nabla^{2}\rho_{B},\nonumber \\
\label{eq:U_a_2_pde}
\end{eqnarray}
where $U_{a}^{(2)}$ is the auxiliary potential under the second-order
approximation. The structure of this equation suggests the decomposition
\begin{equation}
U_{a}^{(2)}(\boldsymbol{r},\boldsymbol{\lambda},\dot{\boldsymbol{\lambda}})=u(\boldsymbol{r},\boldsymbol{\lambda})+\dot{\boldsymbol{\lambda}}\cdot\boldsymbol{\omega}(\boldsymbol{r},\boldsymbol{\lambda}),\label{eq:U_a_2_form}
\end{equation}
with $u$ and $\boldsymbol{\omega}$ satisfying the following constraints:
\begin{equation}
D_{a}\tau_{p}\frac{\partial}{\partial r_{j}^{\beta}}\left(\frac{\partial^{2}(U_{o}+u)}{\partial r_{i}^{\alpha}\partial r_{j}^{\beta}}\rho_{B}\right)=\frac{\partial(U_{o}+u)}{\partial r_{i}^{\alpha}}\rho_{B}+(1+D_{a})\frac{\partial\rho_{B}}{\partial r_{i}^{\alpha}}\label{eq:u_pde}
\end{equation}
and
\begin{equation}
\frac{\partial\rho_{B}}{\partial\lambda_{\mu}}=\frac{\partial}{\partial r_{i}^{\alpha}}\left(\frac{\partial\omega_{\mu}}{\partial r_{i}^{\alpha}}\rho_{B}\right)-D_{a}\tau_{p}\frac{\partial^{2}}{\partial r_{i}^{\alpha}\partial r_{j}^{\beta}}\left(\frac{\partial^{2}\omega_{\mu}}{\partial r_{i}^{\alpha}\partial r_{j}^{\beta}}\rho_{B}\right).\label{eq:omega_pde}
\end{equation}
Further simplification yields the coupled partial differential equations
that determine the auxiliary control under the second-order approximation:
\begin{align}
D_{u}^{(2)}(u) & \equiv\frac{\partial u}{\partial r_{i}^{\alpha}}-D_{a}\tau_{p}\frac{\partial}{\partial r_{j}^{\beta}}\frac{\partial^{2}(U_{o}+u)}{\partial r_{i}^{\alpha}\partial r_{j}^{\beta}}\nonumber \\
 & +\frac{D_{a}\tau_{p}}{1+D_{a}}\frac{\partial^{2}(U_{o}+u)}{\partial r_{i}^{\alpha}\partial r_{j}^{\beta}}\frac{\partial U_{o}}{\partial r_{j}^{\beta}}=0\label{eq:u_pde_1}
\end{align}
and
\begin{align}
D_{\mu}^{(2)}(w_{\mu}) & \equiv(1+D_{a})\nabla^{2}\omega_{\mu}-\nabla_{i}\omega_{\mu}\cdot\nabla_{i}U_{o}+\frac{\partial U_{o}}{\partial\lambda_{\mu}}-\frac{\partial F}{\partial\lambda_{\mu}}\nonumber \\
 & -(1+D_{a})D_{a}\tau_{p}\nabla^{4}\omega_{\mu}+2D_{a}\tau_{p}\nabla_{i}\nabla^{2}\omega_{\mu}\cdot\nabla_{i}U_{o}\nonumber \\
 & -D_{a}\tau_{p}\frac{\partial^{2}\omega_{\mu}}{\partial r_{i}^{\alpha}\partial r_{j}^{\beta}}\left[\frac{1}{1+D_{a}}\frac{\partial U_{o}}{\partial r_{i}^{\alpha}}\frac{\partial U_{o}}{\partial r_{j}^{\beta}}-\frac{\partial^{2}U_{o}}{\partial r_{i}^{\alpha}\partial r_{j}^{\beta}}\right]\nonumber \\
 & =0.\label{eq:omega_pde_1}
\end{align}
Thus, the auxiliary control is fully characterized by these coupled
equations at first and second order.

\section{Approximate shortcut for active matter\label{sec:Approximate-shortcut}}

The above partial differential equations are high-dimensional and
generally intractable for complex potentials. Traditional grid-based
partial differential equation solvers, such as finite difference or
finite element methods, face a fundamental limitation: their computational
cost scales exponentially with dimensionality, rendering them impractical
for high-dimensional systems \cite{bellman1957dynamic,bellman2015adaptive}.
To overcome this limitation, we develop a variational method in this
section. This approach combines parameterized ansatz expressions for
the auxiliary potential \cite{VCD_1,VCD_2,VCD_3} with efficient sampling
technique to determine the approximate potential numerically.

\subsection{Variational formulation (first order)}

Under the first-order approximation, and inspired by the method in
\cite{PhysRevE.103.032146}, we define a set of functionals:

\begin{equation}
G_{\mu}^{(1)}(f_{\mu})\equiv\langle[D_{\mu}^{(1)}(f_{\mu})]^{2}\rangle\label{eq:functional_f}
\end{equation}
for which stationary conditions $\delta G_{\mu}^{(1)}(f_{\mu})/\delta f_{\mu}=0$
are equivalent to the original partial differential equations, Eq.
(\ref{eq:f_pde}). Here, the average $\langle\cdot\rangle$ is defined
as $\langle A\rangle\equiv\int A(\boldsymbol{r})\rho_{B}(\boldsymbol{r},\boldsymbol{\lambda})\text{d}\boldsymbol{r}$.

Unlike previous approaches that employ a single functional \cite{PhysRevE.103.032146},
we define one functional for each component of the control parameter
vector $\boldsymbol{\lambda}$. This one-to-one correspondence decouples
the variational problems for the components of $\boldsymbol{f}$,
and removes explicit dependence on $\dot{\boldsymbol{\lambda}}$,
significantly simplifying the variational procedure.

The key step is to choose an appropriate ansatz form for each $f_{\mu}(\boldsymbol{r},\boldsymbol{\lambda})$.
This selection is guided by physical considerations and the nature
of the experimental platform. Ideally, the ansatz should be as simple
as possible while remaining expressive enough to capture the essential
physics. The related averages $\langle\cdot\rangle$ required in Eq.
(\ref{eq:functional_f}) are evaluated by sampling configurations
generated from simulating the stochastic Langevin dynamics, which
asymptotically converges to $\rho_{B}$. We emphasize that the specific
sampling method is not crucial; alternative approaches such as Monte
Carlo techniques \cite{MC} could be employed equivalently.

Compared with solving Eq. (\ref{eq:f_pde}) using grid-based partial
differential equation solvers, such sampling-based methods approximate
the solution using a finite set of sample points, thereby avoiding
a full grid discretization of the high-dimensional space. This transforms
the computational complexity from an exponential dependence on dimension
(the curse of dimensionality) to a cost that is only weakly dependent
on the dimension.

\subsection{Second-order extension}

For the second-order approximation, we apply the same strategy to
$u$ and $\boldsymbol{\omega}$, defining
\begin{equation}
G_{u}^{(2)}(u)\equiv\langle[D_{u}^{(2)}(u)]^{2}\rangle\label{eq:functional_u}
\end{equation}
and 
\begin{equation}
G_{\mu}^{(2)}(\omega_{\mu})\equiv\langle[D_{\mu}^{(2)}(\omega_{\mu})]^{2}\rangle.\label{eq:functional_omega}
\end{equation}
The stationary conditions recover the governing equations for $u$
and $\boldsymbol{\omega}$. The subsequent procedure---selecting
ansatzes, evaluating related averages via sampling, performing the
variation---follows directly from the first-order case.

\section{The energy cost in swift state transitions\label{sec:energy-cost}}

Within the framework of stochastic thermodynamics \cite{Jarzynski1997,Sekimoto1997},
the mean input work $W\equiv\int dt\langle\partial U/\partial t\rangle$
can be used to quantify the energy cost in a non-equilibrium process.
In such a state-transition process, the energy cost can be expressed
as:

\begin{align}
W & =\Delta\langle U\rangle+\int_{0}^{\tau}\text{d}t\int d\boldsymbol{r}\frac{\boldsymbol{J}_{i}^{2}}{\rho}-(1+D_{a})\Delta S\nonumber \\
 & -D_{a}\tau_{p}\int_{0}^{\tau}\text{d}t\int\text{d}\boldsymbol{r}\frac{J_{i}^{\alpha}}{\rho}\frac{\partial}{\partial r_{j}^{\beta}}\left(\frac{\partial^{2}U}{\partial r_{i}^{\alpha}\partial r_{j}^{\beta}}\rho\right)+\mathcal{O}(\tau_{p}^{3}),\label{eq:work_expression}
\end{align}
where $U\equiv U_{o}+U_{a}$ denotes the total potential and $S\equiv-\int d\boldsymbol{r}\rho\log\rho$
is the Gibbs entropy. The full derivation of Eq. (\ref{eq:work_expression})
is provided in Supplementary Material \cite{sm}.

\subsection{First-order approximation}

Under the first-order approximation, we keep only terms up to first
order in $\tau_{p}$ and the mean input work simplifies to:
\begin{eqnarray}
W^{(1)} & = & \Delta\langle U\rangle-(1+D_{a})\Delta S+\int_{0}^{\tau}\text{d}t\int\text{d}\boldsymbol{r}\frac{(J_{i}^{\alpha})^{2}}{\rho},\nonumber \\
\label{eq:work_1_expression}
\end{eqnarray}
where the probability current is:
\begin{align}
J_{i}^{\alpha} & =-\left[\frac{\partial U}{\partial r_{i}^{\alpha}}\rho+(1+D_{a})\frac{\partial\rho}{\partial r_{i\alpha}^{\alpha}}\right]\nonumber \\
 & =-\frac{\partial(U_{o}+U_{a}^{(1)})}{\partial r_{i}^{\alpha}}\rho+\frac{\partial U_{o}}{\partial r_{i}^{\alpha}}\rho\nonumber \\
 & =-\frac{\partial U_{a}^{(1)}}{\partial r_{i}^{\alpha}}\rho.\label{eq:first-order-current}
\end{align}
The first two terms in Eq. (\ref{eq:work_1_expression}) depend only
on the boundary distributions $\rho_{i}$ and $\rho_{f}$, and thus
protocol-independent. The remaining contribution defines the dissipative
work, and denoted by $W_{\text{d}}^{(1)}$, must be minimized. Substitute
Eq. (\ref{eq:U_a_1_form}) and $\rho(\boldsymbol{r},\boldsymbol{\lambda})=\rho_{B}(\boldsymbol{r},\boldsymbol{\lambda})$
into Eq. (\ref{eq:first-order-current}), the dissipative work can
be recast in a geometric form:
\begin{equation}
W_{\text{d}}^{(1)}=\int_{0}^{\tau}\text{d}t\dot{\lambda}_{\mu}\dot{\lambda}_{\nu}g_{\mu\nu}^{(1)},\label{eq:work_1_expression_2}
\end{equation}
where the metric tensor 
\begin{equation}
g_{\mu\nu}^{(1)}=\langle\left(\nabla_{i}f_{\mu}\right)\cdot\left(\nabla_{i}f_{\nu}\right)\rangle\label{eq:metric_first}
\end{equation}
 is positive semi-definite. A detailed derivation of Eqs. (\ref{eq:work_1_expression_2})
and (\ref{eq:metric_first}) is provided in Supplementary Material
\cite{sm}.

\subsection{Second-order approximation}

Under the second-order approximation, we keep only terms up to second
order in $\tau_{p}$ and the mean input work becomes:
\begin{eqnarray}
W_{\text{ }}^{(2)} & = & \Delta\langle U\rangle-(1+D_{a})\Delta S+\int_{0}^{\tau}\text{d}t\int\text{d}\boldsymbol{r}\frac{(J_{i}^{\alpha})^{2}}{\rho}\nonumber \\
 &  & -D_{a}\tau_{p}\int_{0}^{\tau}\text{d}t\int\text{d}\boldsymbol{r}\frac{J_{i}^{\alpha}}{\rho}\frac{\partial}{\partial r_{j}^{\beta}}\left(\frac{\partial^{2}U}{\partial r_{i}^{\alpha}\partial r_{j}^{\beta}}\rho\right).\label{eq:work_2_expreesion}
\end{eqnarray}
Using the structure of Eq. (\ref{eq:omega_pde}), the current can
be written as 
\begin{equation}
J_{i}^{\alpha}=\dot{\lambda}_{\mu}h_{i\mu}^{\alpha}\label{eq:second-order-current}
\end{equation}
with
\begin{equation}
h_{i\mu}^{\alpha}=-\left[\frac{\partial\omega_{\mu}}{\partial r_{i}^{\alpha}}\rho+D_{a}\tau_{p}\frac{\partial}{\partial r_{j}^{\beta}}\left(\frac{\partial^{2}\omega_{\mu}}{\partial r_{i}^{\alpha}\partial r_{j}^{\beta}}\rho\right)\right].
\end{equation}
Substitute Eqs. (\ref{eq:U_a_2_form}), (\ref{eq:second-order-current}),
and $\rho(\boldsymbol{r},\boldsymbol{\lambda})=\rho_{B}(\boldsymbol{r},\boldsymbol{\lambda})$
into Eq. (\ref{eq:work_2_expreesion}), the mean input work can be
recast as
\begin{align}
W_{\text{ }}^{(2)} & =\Delta\langle U\rangle-\int_{0}^{\tau}\text{d}t\int\text{d}\boldsymbol{r}\left[\frac{\partial\rho_{B}}{\partial t}(U_{o}+u)\right]\nonumber \\
 & +\int_{0}^{\tau}\text{d}t\dot{\lambda}_{\mu}\dot{\lambda}_{\nu}g_{\mu\nu}^{(2)},\label{eq:work_2_expreesion_2}
\end{align}
where the metric tensor 
\begin{equation}
g_{\mu\nu}^{(2)}=\langle\left(\nabla_{i}\omega_{\mu}\right)\cdot\left(\nabla_{i}\omega_{\nu}\right)\rangle+D_{a}\tau_{p}\langle\left(\nabla_{i}\nabla_{j}\omega_{\mu}\right)\cdot\left(\nabla_{i}\nabla_{j}\omega_{\nu}\right)\rangle\label{eq:metric_second}
\end{equation}
 is also positive semi-definite. A detailed derivation of Eqs. (\ref{eq:work_2_expreesion_2})
and (\ref{eq:metric_second}) is provided in Supplementary Material
\cite{sm}.

As for the second term in Eq. (\ref{eq:work_2_expreesion_2}), note
that both $U_{o}$ and $u$ can be viewed as functionals of $\rho$.
Hence, there exists a functional $H[\rho]$ such that $\delta H[\rho]/\delta\rho=U_{o}+u$.
Consequently, this contribution reduces to a boundary term and is
therefore protocol-independent. The protocol-dependent dissipative
work therefore takes a geometric form:
\begin{equation}
W_{\text{d}}^{(2)}=\int_{0}^{\tau}\text{d}t\dot{\lambda}_{\mu}\dot{\lambda}_{\nu}g_{\mu\nu}^{(2)}.
\end{equation}

\subsection{Geometric structure and optimal protocols}

In both the first- and second-order approximations\textbf{, $\boldsymbol{g}$
}defines a positive semi-definite metric, referred to as the thermodynamic
metric \cite{MeasuringThermodynamicLength,Thermodynamic_Metrics_and_OptimalPaths,Geometry_Li}.
In the Riemannian manifold induced by the thermodynamic metric, the
dissipative work $W_{\text{d}}$ is bounded from below by $\frac{\mathcal{L}^{2}}{\tau}$.
Here, $\mathcal{L}\equiv\int_{0}^{\tau}dt\sqrt{\dot{\lambda}_{\mu}\dot{\lambda}_{\nu}g_{\mu\nu}}$
is the thermodynamic length. This length is that of the geodesic path
in the geometric space described by the thermodynamic metric $\boldsymbol{g}$.
The geodesic path represents the optimal protocol, which can be found
by solving the geodesic equations.

To summarize, the proposed framework consists of three key steps---shortcut
design, auxiliary control construction, and dissipation minimization---as
outlined in Table \ref{tab:short_cut_framework}. 
\begin{table*}
\begin{centering}
\begin{tabular}{ll}
\hline 
Stage & Core Steps\tabularnewline
\hline 
$\;\;$ & $\;\;$\tabularnewline
I. Shortcut design & \textbullet{} Design an evolutionary path $\rho_{B}(\boldsymbol{r},\boldsymbol{\lambda}(t))$
connecting $\rho_{i}$ and $\rho_{f}$\tabularnewline
 & \textbullet{} Steer the system evolve along $\rho_{B}$ by introducing
$U_{a}$ into evolution equation\tabularnewline
$\;\;$ & $\;\;$\tabularnewline
II. Auxiliary control & Control equations:\tabularnewline
 & $\;\;\;\;$\textbullet{} First\nobreakdash-order: $U_{a}^{(1)}(\boldsymbol{r},\boldsymbol{\lambda},\dot{\boldsymbol{\lambda}})=\dot{\boldsymbol{\lambda}}\cdot\boldsymbol{f}(\boldsymbol{r},\boldsymbol{\lambda})$,
$D_{\mu}^{(1)}(f_{\mu})=0$\tabularnewline
 & $\;\;\;\;$\textbullet{} Second\nobreakdash-order: $U_{a}^{(2)}(\boldsymbol{r},\boldsymbol{\lambda},\dot{\boldsymbol{\lambda}})=u(\boldsymbol{r},\boldsymbol{\lambda})+\dot{\boldsymbol{\lambda}}\cdot\boldsymbol{\omega}(\boldsymbol{r},\boldsymbol{\lambda})$,
$D_{u}^{(2)}(u)=0$, $D_{\mu}^{(2)}(w_{\mu})=0$\tabularnewline
 & Solution:\tabularnewline
 & $\;\;\;\;$\textbullet{} If analytically solvable \textrightarrow{}
exact $\boldsymbol{f}(\boldsymbol{r},\boldsymbol{\lambda})$, $u(\boldsymbol{r},\boldsymbol{\lambda})$,
$\boldsymbol{\omega}(\boldsymbol{r},\boldsymbol{\lambda})$\tabularnewline
 & $\;\;\;\;$\textbullet{} Otherwise (complex interactions):\tabularnewline
 & $\;\;\;\;$$\;\;\;\;$-- Variational functionals $G=\langle[D(\cdot)]^{2}\rangle$
with parameterized ansatzes\tabularnewline
 & $\;\;\;\;$$\;\;\;\;$-- Sample from $\rho_{B}$, determine coefficients
in ansatzes on discrete $\boldsymbol{\lambda}$\nobreakdash-grid\tabularnewline
 & $\;\;\;\;$$\;\;\;\;$-- Interpolate to obtain continuous functions\tabularnewline
$\;\;$ & $\;\;$\tabularnewline
III. Minimizing dissipation & \textbullet{} Dissipative work: $W_{\text{d}}=\int_{0}^{\tau}\text{d}t\dot{\lambda}_{\mu}\dot{\lambda}_{\nu}g_{\mu\nu}$,
with metric $\boldsymbol{g}$ defined from $\boldsymbol{f}$ or $\boldsymbol{\omega}$\tabularnewline
 & \textbullet{} Optimal protocol = geodesic\tabularnewline
$\;\;$ & $\;\;$\tabularnewline
\hline 
 & \tabularnewline
\hline 
\end{tabular}
\par\end{centering}
\caption{\label{tab:short_cut_framework}Main steps for realizing swift state
transitions in active systems.}
\end{table*}

\begin{figure*}
\begin{centering}
\includegraphics[width=0.8\paperwidth,keepaspectratio,height=0.8\paperwidth]{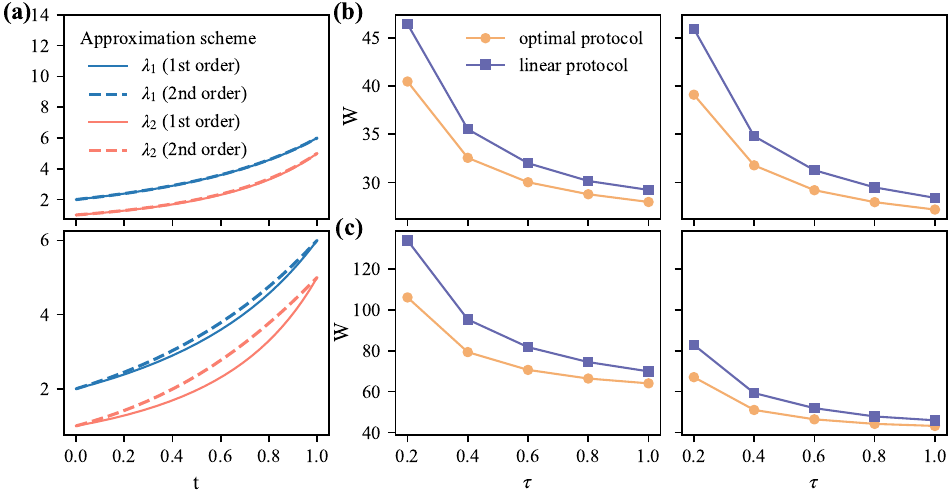}
\par\end{centering}
\caption{\label{fig:coupling_to_mass_center}Optimal control and mean input
work in an active system of particles harmonically coupled to their
center of mass.\textbf{ (a)} Geodesics (optimal protocols) obtained
from the thermodynamic metrics $\boldsymbol{g}^{(1)}$ (dashed) and
$\boldsymbol{g}^{(2)}$ (solid). For fixed activity parameters, the
geodesic shape is independent of the process duration $\tau$; results
are shown for a reference value $\tau=1.0$. \emph{Top panel}: $D_{a}=1$,
$\tau_{p}=0.01$. \emph{Bottom panel}: $D_{a}=5$, $\tau_{p}=0.1$.
\textbf{(b)} Mean input work as a function of the process duration
$\tau$, evaluated along the geodesics shown in the top panel of \textbf{(a)}
and compared with a linear protocol. \emph{Left panel}: work for the
$\boldsymbol{g}^{(1)}$ geodesic. \emph{Right panel}: work for the
$\boldsymbol{g}^{(2)}$ geodesic.\textbf{ (c)} Same as \textbf{(b)},
but for the geodesic shown in the bottom panel of \textbf{(a)}.}
\end{figure*}

\section{Applications}

In this section, we validate the proposed framework through two representative
examples: an analytically solvable system with attractive harmonic
coupling, and a numerically tractable system with repulsive Gaussian-core
interactions. These two cases are complementary in both interaction
type (attractive vs. repulsive) and methodology (analytic vs. numerical),
thereby demonstrating the generality and robustness of our approach.

\subsection{Harmonic Coupling\label{subsec:harmonic_coupling}}

We begin by considering an active system confined in a harmonic trap
with time-dependent strength $\lambda_{1}(t)$. In addition to this
external confinement, each particle $i$ is coupled to the center
of mass of the system, $\boldsymbol{R}\equiv\frac{1}{N}\sum_{i}\boldsymbol{r}_{i}$,
via the harmonic coupling potential $\frac{1}{2}\lambda_{2}(t)(\boldsymbol{r}_{i}-\boldsymbol{R})^{2}$.
This interaction is simple yet captures the essential character that
some active particles are sensitive to their distance from the group
\cite{harmonic_interaction_1,harmonic_interaction_2,harmonic_interaction_3,harmonic_interaction_4,harmonic_interaction_5}.
The total potential of the system is therefore given by:
\begin{equation}
U_{o}=\frac{1}{2}\lambda_{1}(t)\boldsymbol{r}_{i}^{2}+\frac{1}{2}\lambda_{2}(t)(\boldsymbol{r}_{i}-\boldsymbol{R})^{2}.
\end{equation}
Let us introduce the coordinate vector for the $\alpha$-th spatial
component as $\boldsymbol{r}^{\alpha}\equiv\left(r_{1}^{\alpha},r_{2}^{\alpha},\dots,r_{N}^{\alpha}\right)^{T}$.
In this notation, the potential can be expressed compactly in a quadratic
form,
\begin{equation}
U_{o}=\frac{1}{2}\left(\boldsymbol{r}^{\alpha}\right)^{\text{T}}\mathbf{\Lambda}\boldsymbol{r}^{\alpha},\label{eq:harmonic_U_o}
\end{equation}
where the coupling matrix $\mathbf{\Lambda}$ is defined as $\mathbf{\Lambda}\equiv\lambda_{1}\mathbf{I}+\lambda_{2}\mathbf{C}$.
Here, $\mathbf{I}$ is the $N\times N$ identity matrix representing
the external harmonic trap, and $\mathbf{C}$ is the centering matrix
that encodes the interactions relative to the center of mass. Full
details of the derivation Eq.(\ref{eq:harmonic_U_o}) are provided
in Supplementary Material \cite{sm}.

In the limit of short persistence time, we neglect high-order terms
of the expansion of the last term in the probability current, which
allows us to derive an exact solution for the auxiliary potential
$U_{a}$ that guides the probability distribution to evolve alongside
$\rho_{B}$:
\begin{equation}
U_{a}^{(1)}=\frac{1}{4}\left(\boldsymbol{r}^{\alpha}\right)^{\text{T}}\mathbf{\Lambda}^{-1}\dot{\mathbf{\Lambda}}\boldsymbol{r}^{\alpha}
\end{equation}
and

\begin{equation}
U_{a}^{(2)}=\frac{1}{2}\left(\boldsymbol{r}^{\alpha}\right)^{\text{T}}\left(\boldsymbol{I}+\frac{D_{a}\tau_{p}}{1+D_{a}}\mathbf{\Lambda}\right)^{-1}\left(\mathbf{\Lambda}+\frac{1}{2}\mathbf{\Lambda}^{-1}\dot{\mathbf{\Lambda}}\right)\boldsymbol{r}^{\alpha}.
\end{equation}
From these expressions, the functions $f_{\mu}$ and $\omega_{\mu}$
can be identified, and the corresponding thermodynamic metrics $\boldsymbol{g}^{(1)}$
and $\boldsymbol{g}^{(2)}$ follow (see \cite{sm}). The optimal protocols
are then obtained by solving the associated geodesic equations.

Figure \ref{fig:coupling_to_mass_center} illustrates the geodesic
profiles and the corresponding mean input work $W$ for selected activity
parameters. As illustrated in Fig. \ref{fig:coupling_to_mass_center}a,
for small activity parameters (e.g., $D_{a}=1,\tau_{p}=0.01$), the
geodesics obtained from the first- and second-order approximation
schemes almost overlap, as the short persistence time $\tau_{p}$
makes the second-order corrections negligible. In contrast, under
stronger activity (e.g., $D_{a}=5,\tau_{p}=0.1$), the second-order
term contributes significantly, leading to a clear separation between
the geodesics from the two approximation schemes. Figures \ref{fig:coupling_to_mass_center}b
and \ref{fig:coupling_to_mass_center}c show $W$ as a function of
the process duration $\tau$ for both the geodesic protocols and reference
linear protocols. The geodesic protocols consistently exhibit lower
dissipation than the linear protocols across all durations, confirming
its optimality within the geometric framework. When activity parameters
are small, the first- and second-order approximation schemes perform
similarly, as expected. For larger activity parameters, however, the
second-order approximation scheme not only consumes less work but
also achieves more accurate state transition.

\subsection{Pairwise Gaussian-core Interaction\label{subsec:gaussian_coupling}}

While the above example admits an analytic solution, such closed-form
results are often unavailable for more complex interaction. To overcome
this limitation, we now resort to the numerical methods. As in the
previous section, we include an external harmonic trap with time-dependent
strength $\lambda_{1}(t)$. The pairwise interaction potential between
particles $i$ and $j$ is given by a Gaussian form: $\lambda_{2}(t)\exp(-\frac{r_{ij}^{2}}{2\sigma^{2}})$,
where $r_{ij}$ is the interparticle distance. This potential is characteristic
of a Gaussian-core fluid, a classical model for soft, penetrable particles
\cite{gaussian-core_1,gaussian-core_2}. Here, the strength parameter
$\lambda_{2}>0$ is treated as a tunable, time-dependent variable,
while $\sigma$ defines the effective soft radius of a particle. The
total potential energy of the system is therefore:
\begin{equation}
U_{o}=\frac{1}{2}\lambda_{1}(t)\boldsymbol{r}_{i}^{2}+\frac{1}{2}\lambda_{2}(t)\sum_{i\neq j}\exp(-\frac{r_{ij}^{2}}{2\sigma^{2}}).\label{eq:gaussian-core_U_o}
\end{equation}

\begin{figure*}
\centering{}\includegraphics[width=0.8\paperwidth,keepaspectratio,height=0.8\paperwidth]{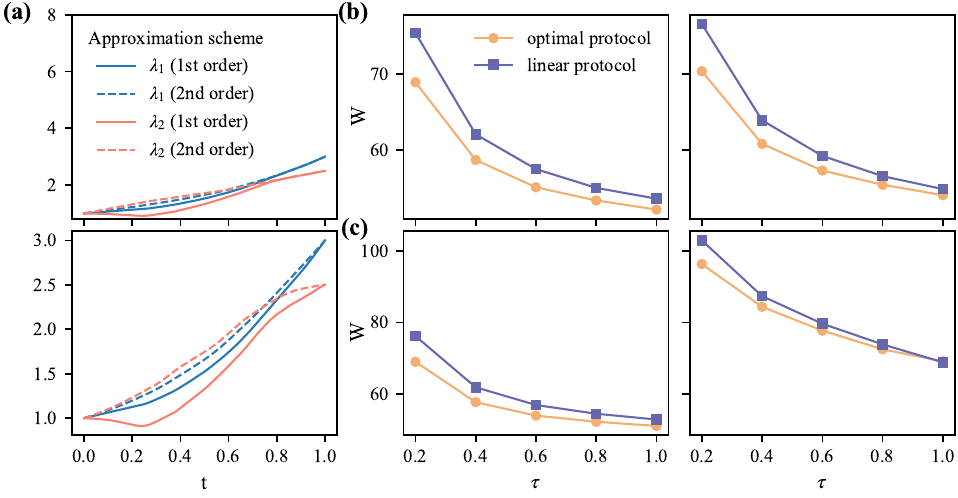}\caption{\label{fig:Gaussian-core}Optimal control and mean input work in an
active system with Gaussian-core pair interactions.\textbf{ (a)} Geodesics
(optimal protocols) obtained from the thermodynamic metrics $\boldsymbol{g}^{(1)}$
(dashed) and $\boldsymbol{g}^{(2)}$ (solid). For fixed activity parameters,
the geodesic shape is independent of the process duration $\tau$;
Results are shown for a reference value $\tau=1.0$. \emph{Top panel}:
$D_{a}=1$, $\tau_{p}=0.01$. \emph{Bottom panel}: $D_{a}=1$, $\tau_{p}=0.1$.
\textbf{(b)} Mean input work as a function of process duration $\tau$,
evaluated along the geodesics shown in the top panel of \textbf{(a)}
and compared with a linear protocol. \emph{Left panel}: work for the
$\boldsymbol{g}^{(1)}$ geodesic. \emph{Right panel}: work for the
$\boldsymbol{g}^{(2)}$ geodesic.\textbf{ (c)} Same as \textbf{(b)},
but for the geodesic shown in the bottom panel of \textbf{(a)}.}
\end{figure*}

Under the first-order approximation, we adopt the following ansatzes
for $f_{1}$ and $f_{2}$ :
\begin{eqnarray}
f_{1}(\boldsymbol{r},\boldsymbol{\lambda}) & = & a(\boldsymbol{\lambda})\sum_{i\neq j}r_{ij}^{2}+b(\boldsymbol{\lambda})\boldsymbol{r}_{i}^{2}\label{eq:ansatz_1_first}
\end{eqnarray}
and

\begin{equation}
f_{2}(\boldsymbol{r},\boldsymbol{\lambda})=c(\boldsymbol{\lambda})\sum_{i\neq j}r_{ij}^{2}+d(\boldsymbol{\lambda})\boldsymbol{r}_{i}^{2},\label{eq:ansatz_2_first}
\end{equation}
where $a,b,c,d$ are coefficients to be determined variationally.
The ansatzes both consist of two distinct components: one depends
on the interparticle distance to account for the lag arising from
internal interactions, and the other depends on the absolute positions
to account for the lag induced by the external potential. As for the
specific forms, we adopt the simplest forms in Eq. (\ref{eq:ansatz_1_first})
and Eq. (\ref{eq:ansatz_2_first}), as it suffices for our purpose.
More complex forms may offer greater precision, but they do not change
the procedure. To eliminate terms involving the normalization factor
$F$ in Eq. (\ref{eq:functional_f}) without affecting the variational
result, we discard irrelevant terms with $f_{\mu}$ and boundary terms
after integration by parts (see details in Supplementary Material
\cite{sm}). This yields the simplified functional:
\begin{eqnarray}
G_{\mu}(f_{\mu}) & = & \langle\left[(1+D_{a})\nabla^{2}f_{\mu}-\nabla_{i}f_{\mu}\cdot\nabla_{i}U_{o}\right]^{2}\label{eq:functional_f_2}\\
 &  & -2(1+D_{a})\nabla_{i}f_{\mu}\cdot\nabla_{i}\frac{\partial U_{o}}{\partial\lambda_{\mu}}\rangle\text{ }(\text{no sum on \ensuremath{\mu}}).\nonumber 
\end{eqnarray}

Substituting Eqs. (\ref{eq:ansatz_1_first}) and (\ref{eq:ansatz_2_first})
into Eq. (\ref{eq:functional_f_2}) and performing the variational
procedure yields a system of linear equations for $a,b,c,d$. The
coefficients in this linear system are expressed as ensemble averages,
which can be evaluated efficiently by direct sampling from $\rho_{B}$
at the given $\boldsymbol{\lambda}$. Solving this system gives the
explicit forms of $f_{1}$ and $f_{2}$, from which the corresponding
thermodynamic metric at fixed $\left(\lambda_{1},\lambda_{2}\right)$
can be also derived. The computation is executed over a discrete grid
that spans the parameter region of interest, producing a numerical,
pointwise representation of both the auxiliary control and the metric.
The continuous functions are subsequently constructed via cubic spline
interpolation of this discrete dataset. Finally, we numerically solve
the geodesic equation on this interpolated metric to obtain the optimal
(geodesic) protocol trajectory.

Under the second-order approximation, we adopt the following ansatzes
for $u$, $\omega_{1}$ and $\omega_{2}$:
\begin{equation}
u=a(\boldsymbol{\lambda})\sum_{i\neq j}\exp(-\frac{r_{ij}^{2}}{2\sigma^{2}})+b(\boldsymbol{\lambda})\boldsymbol{r}_{i}^{2},
\end{equation}
\begin{equation}
\omega_{1}=c(\boldsymbol{\lambda})\sum_{i\neq j}r_{ij}^{2}+d(\boldsymbol{\lambda})\boldsymbol{r}_{i}^{2}
\end{equation}
and

\begin{equation}
\omega_{2}=e(\boldsymbol{\lambda})\sum_{i\neq j}r_{ij}^{2}+f(\boldsymbol{\lambda})\boldsymbol{r}_{i}^{2}.
\end{equation}
The component $u$ is part of $U_{a}^{(2)}$ but does not contribute
to \textbf{$\boldsymbol{g}^{(2)}$}, whereas $\omega_{1}$ and $\omega_{2}$
are the variables that generate \textbf{$\boldsymbol{g}^{(2)}$}.
Here, $u$ serves as a correction to Eq.(\ref{eq:gaussian-core_U_o})
and is designed to have a similar functional form to facilitate experimental
control. Note that $\partial F/\partial\lambda_{\mu}=\langle\partial U_{o}/\partial\lambda_{\mu}\rangle$
(see \cite{sm}) should be inserted into Eq. (\ref{eq:functional_omega})
since $F$ is unknown. Following a similar procedure, we can obtain
numerical representations of $U_{a}^{(2)}$ and $\boldsymbol{g}^{(2)}$,
from which the geodesics are then determined.

Related results are shown in Fig. \ref{fig:Gaussian-core} which are
similar to the main trends observed in Fig. \ref{fig:coupling_to_mass_center}.
However, it is worth noting that in Fig. \ref{fig:Gaussian-core}c
the second-order approximation scheme consumes more work than the
first-order scheme, unlike in Fig. \ref{fig:coupling_to_mass_center}c.
This difference stems from the use of a repulsive interaction in the
current section, as opposed to the attractive interaction considered
previously. Another observation is that in the right panel, as the
process duration increases, the optimal protocol gradually approaches
its linear counterpart. We attribute this behavior to the limited
expressiveness of the chosen ansatzes used for $U_{a}^{(2)}$. Employing
a more flexible parameterization---such as one based on neural networks---could
potentially yield better results \cite{xu2022reinforcement}.

\section{Conclusions}

In this work, we have developed a general shortcut framework for realizing
swift state transitions in active systems. Working in the weak activity
regime, this framework retains terms up to first order and second
order in $\tau_{p}$. By introducing an auxiliary potential, we can
guide the system along a predefined distribution path, enabling it
to reach target states in finite time---a task that would otherwise
require infinitely long driving under original dynamics. This approach
extends the concept of shortcuts from passive systems to active systems.
A key challenge in such finite-time processes is the unavoidable energy
cost. To address this, we have employed thermodynamic geometry to
quantify and minimize the dissipative work. By expressing the dissipative
work in a geometric form, we have identified a positive semi-definite
thermodynamic metric on the space of control parameters. Within this
Riemannian manifold, the optimal protocol that minimizes dissipation
corresponds to the geodesic path. This geometric perspective provides
a systematic and elegant framework for protocol optimization in active
systems. We have demonstrated the applicability and robustness of
our approach through two complementary case studies. For an attractive
harmonically coupled system, we have derived exact analytical expressions
for the auxiliary potential and thermodynamic metric under both the
first- and second-order approximations. This solvable model has validated
the core ideas and has revealed how increasing activity parameters
lead to noticeable deviations between approximation orders. For the
more complex repulsive Gaussian-core pairwise interaction, where analytical
solutions are intractable, we have developed a variational method
that combines parameterized ansatzes with sampling technique. This
numerical approach overcoming the curse of dimensionality inherent
in grid-based methods.

Our results consistently show that geodesic protocols derived from
the thermodynamic metric outperform naive linear protocols in reducing
dissipation across all process durations. Interestingly, compared
to the first-order approximations, the second-order schemes not only
ensure more accurate state transformations but also yield lower work
consumption for attractive interparticle interactions, while resulting
in higher work consumption for repulsive interactions. The contrasting
performance of the second-order approximation schemes for attractive
and repulsive interactions reveals its ability to capture their distinct
physical mechanisms.

Several extensions are worth future investigation. First, the present
framework is restricted to weak activity; extending it to strong activity
using non-perturbative treatments of the Fox approximation would be
highly valuable. Second, the expressiveness of our ansatzes could
be enhanced by employing more flexible parameterizations, such as
neural networks \cite{casert2024learning,whitelam2025benchmark} or
linear spatiotemporal basis parametrizations \cite{zhong2024time},
potentially yielding higher accuracy and broader applicability to
more complex interactions.

\bibliographystyle{apsrev4-1}
\bibliography{main_bibtex}

\end{document}

\title{Supplementary Material: Shortcuts to state transitions for active
matter}
\author{Guodong Cheng}
\affiliation{School of Physics and Astronomy, Beijing Normal University, Beijing
100875, China}
\author{Z. C. Tu}
\affiliation{School of Physics and Astronomy, Beijing Normal University, Beijing
100875, China}
\author{Geng Li}
\email{gengli@bnu.edu.cn}

\affiliation{School of Systems Science, Beijing Normal University, Beijing 100875,
China}

\maketitle
\tableofcontents{}

\section{Derivation of the probability distribution evolution equation based
on the Fox approximation}

We consider a system of interacting active Ornstein-Uhlenbeck particles
(AOUPs). The dynamics of particle $i$ in spatial direction $\alpha$
is governed by the overdamped Langevin equation: 
\begin{equation}
\dot{r}_{i}^{\alpha}(t)=F_{i}^{\alpha}(\boldsymbol{r})+\xi_{i}^{\alpha}(t)+\eta_{i}^{\alpha}(t).\label{eq:SMAOUP}
\end{equation}
The probability distribution of the system is defined as $\rho(\boldsymbol{r},t)=\langle\delta_{D}\left(\boldsymbol{r}-\boldsymbol{r}(t)\right)\rangle$.
Throughout this section, $\delta_{D}$ denotes the Dirac delta function,
while $\delta$ is reserved for functional variations. The time evolution
of $\rho(\boldsymbol{r},t)$ is governed by

\begin{eqnarray}
\frac{\partial\rho}{\partial t} & = & \frac{\partial}{\partial t}\langle\delta_{D}(\boldsymbol{r}-\boldsymbol{r}(t))\rangle\nonumber \\
 & = & \langle\frac{\partial}{\partial t}\delta_{D}(\boldsymbol{r}-\boldsymbol{r}(t))\rangle\nonumber \\
 & = & \langle\dot{r}_{i}^{\alpha}(t)\frac{\partial\delta_{D}(\boldsymbol{r}-\boldsymbol{r}(t))}{\partial r_{i}^{\alpha}(t)}\rangle\nonumber \\
 & = & \langle(F_{i}^{\alpha}(\boldsymbol{r}(t))+\xi_{i}^{\alpha}(t)+\eta_{i}^{\alpha}(t))\frac{\partial}{\partial r_{i}^{\alpha}(t)}\delta_{D}(\boldsymbol{r}-\boldsymbol{r}(t))\rangle\nonumber \\
 & = & -\langle(F_{i}^{\alpha}(\boldsymbol{r}(t))+\xi_{i}^{\alpha}(t)+\eta_{i}^{\alpha}(t))\frac{\partial}{\partial r_{i}^{\alpha}}\delta_{D}(\boldsymbol{r}-\boldsymbol{r}(t))\rangle\nonumber \\
 & = & -\frac{\partial}{\partial r_{i}^{\alpha}}\langle(F_{i}^{\alpha}(\boldsymbol{r}(t))+\xi_{i}^{\alpha}(t)+\eta_{i}^{\alpha}(t))\delta_{D}(\boldsymbol{r}-\boldsymbol{r}(t))\rangle\nonumber \\
 & = & -\frac{\partial}{\partial r_{i}^{\alpha}}\left(F_{i}^{\alpha}(\boldsymbol{r})\rho(\boldsymbol{r},t)+\langle\xi_{i}^{\alpha}\delta_{D}(\boldsymbol{r}-\boldsymbol{r}(t))\rangle+\langle\eta_{i}^{\alpha}\delta_{D}(\boldsymbol{r}-\boldsymbol{r}(t))\rangle\right).\label{eq:evolution_equation}
\end{eqnarray}
The central task is therefore to evaluate the correlations $\langle\xi_{i}^{\alpha}\delta_{D}(\boldsymbol{r}-\boldsymbol{r}(t))\rangle$
and $\langle\eta_{i}^{\alpha}\delta_{D}(\boldsymbol{r}-\boldsymbol{r}(t))\rangle$.
To this end, we employ the Novikov theorem \citep{novikov1965functionals}:
\begin{equation}
\langle\xi_{i}^{\alpha}(t)Q[\boldsymbol{\xi}]\rangle=\int\text{d}s\langle\xi_{i}^{\alpha}(t)\xi_{j}^{\beta}(s)\rangle\langle\frac{\delta Q[\boldsymbol{\xi}]}{\delta\xi_{j}^{\beta}(s)}\rangle,\label{eq:Novikov}
\end{equation}
which is a standard result for Gaussian stochastic processes. Specifically,
for a zero\nobreakdash-mean Gaussian random field $\boldsymbol{\xi}$,
the correlation between $\boldsymbol{\xi}$ and an arbitrary functional
$Q[\boldsymbol{\xi}]$ can be expressed as an integral involving the
noise covariance multiplied by the mean functional derivative of $Q$
with respect to $\boldsymbol{\xi}$.

We begin by evaluating the term $\langle\xi_{i}^{\alpha}(t)\delta_{D}(\boldsymbol{r}-\boldsymbol{r}(t))\rangle$
in Eq. (\ref{eq:evolution_equation}) using Eq. (\ref{eq:Novikov}):
\begin{eqnarray}
\langle\xi_{i}^{\alpha}(t)\delta_{D}(\boldsymbol{r}-\boldsymbol{r}(t))\rangle & = & \int_{-\infty}^{t}\text{d}s\langle\xi_{i}^{\alpha}(t)\xi_{j}^{\beta}(s)\rangle\langle\frac{\delta[\delta_{D}(\boldsymbol{r}-\boldsymbol{r}(t))]}{\delta\xi_{j}^{\beta}(s)}\rangle\nonumber \\
 & = & 2\int_{-\infty}^{t}\text{d}s\delta^{\alpha\beta}\delta_{ij}\delta_{D}(t-s)\langle\frac{\delta[\delta_{D}(\boldsymbol{r}-\boldsymbol{r}(t))]}{\delta\xi_{j}^{\beta}(s)}\rangle\nonumber \\
 & = & 2\int_{-\infty}^{t}\text{d}s\delta^{\alpha\beta}\delta_{ij}\delta_{D}(t-s)\langle\frac{\partial\delta_{D}(\boldsymbol{r}-\boldsymbol{r}(t))}{\partial r_{k}^{\gamma}(t)}\frac{\delta r_{k}^{\gamma}(t)}{\delta\xi_{j}^{\beta}(s)}\rangle\nonumber \\
 & = & -2\int_{-\infty}^{t}\text{d}s\delta^{\alpha\beta}\delta_{ij}\delta_{D}(t-s)\langle\frac{\partial\delta_{D}(\boldsymbol{r}-\boldsymbol{r}(t))}{\partial r_{k}^{\gamma}}\frac{\delta r_{k}^{\gamma}(t)}{\delta\xi_{j}^{\beta}(s)}\rangle\nonumber \\
 & = & -2\int_{-\infty}^{t}\text{d}s\delta^{\alpha\beta}\delta_{ij}\delta_{D}(t-s)\langle\frac{\partial\delta_{D}(\boldsymbol{r}-\boldsymbol{r}(t))}{\partial r_{k}^{\gamma}}\theta(t-s)\delta^{\beta\gamma}\delta_{jk}\rangle\nonumber \\
 & = & -2\delta^{\alpha\beta}\delta_{ij}\frac{\partial}{\partial r_{j}^{\beta}}\int ds\langle\delta_{D}(\boldsymbol{r}-\boldsymbol{r}(t))\rangle\theta(t-s)\delta_{D}(t-s)\nonumber \\
 & = & -\delta^{\alpha\beta}\delta_{ij}\frac{\partial}{\partial r_{j}^{\beta}}\rho(\boldsymbol{r},t)\nonumber \\
 & = & -\nabla_{j}^{\beta}\rho(\boldsymbol{r},t).\label{eq:term_2}
\end{eqnarray}
In the second line, we have used the correlation identity $\langle\xi_{i}^{\alpha}(t)\xi_{j}^{\beta}(s)\rangle=2\delta^{\alpha\beta}\delta_{ij}\delta(t-s)$.
In the fifth line, we have employed the identity 
\begin{equation}
\frac{\delta r_{k}^{\gamma}(t)}{\delta\xi_{j}^{\beta}(s)}=\theta(t-s)\delta^{\beta\gamma}\delta_{jk},\label{eq:white_variation}
\end{equation}
which follows from the integral form of the Langevin equation. Specifically,

\begin{equation}
r_{k}^{\gamma}(t)=r_{k}^{\gamma}(0)+\int_{0}^{t}F_{k}^{\gamma}(\boldsymbol{r}(t^{\prime}))\text{d}t^{\prime}+\int_{0}^{t}\xi_{k}^{\gamma}(t^{\prime})\text{d}t^{\prime}+\int_{0}^{t}\eta_{k}^{\gamma}(t^{\prime})\text{d}t^{\prime}.
\end{equation}
Taking the functional derivative of $\boldsymbol{r}(t)$ with respect
to $\xi_{j}^{\beta}(s)$ yields
\begin{eqnarray}
\frac{\delta r_{k}^{\gamma}(t)}{\delta\xi_{j}^{\beta}(s)} & = & \frac{\delta}{\delta\xi_{j}^{\beta}(s)}\int_{0}^{t}F_{k}^{\gamma}(\boldsymbol{r}(t^{\prime}))\text{d}t^{\prime}+\frac{\delta}{\delta\xi_{j}^{\beta}(s)}\int_{0}^{t}\xi_{k}^{\gamma}(t^{\prime})\text{d}t^{\prime}\nonumber \\
 & = & \int_{0}^{t}\frac{\partial F_{k}^{\gamma}(\boldsymbol{r}(t^{\prime}))}{\partial r_{l}^{\sigma}(t^{\prime})}\frac{\delta r_{l}^{\sigma}(t^{\prime})}{\delta\xi_{j}^{\beta}(s)}\text{d}t^{\prime}+\int_{0}^{t}\delta^{\beta\gamma}\delta_{jk}\delta_{D}(t^{\prime}-s)\text{d}t^{\prime}\nonumber \\
 & = & \int_{0}^{t}\frac{\partial F_{k}^{\gamma}(\boldsymbol{r}(t^{\prime}))}{\partial r_{l}^{\sigma}(t^{\prime})}\frac{\delta r_{l}^{\sigma}(t^{\prime})}{\delta\xi_{j}^{\beta}(s)}\text{d}t^{\prime}+\theta(t-s)\delta^{\beta\gamma}\delta_{jk}\nonumber \\
 & = & \theta(t-s)\delta^{\beta\gamma}\delta_{jk}.
\end{eqnarray}
 In the final step, since the correlation time of white noise is zero,
the response $\frac{\delta r_{l}^{\sigma}(t^{\prime})}{\delta\xi_{j}^{\beta}(s)}$
is effectively local,  confining the integral to the correlation time
interval. As this interval vanishes in the white-noise limit, the
contribution is negligible, thereby leading to Eq. (\ref{eq:white_variation}).

We next evaluate the term $\langle\eta_{i}^{\alpha}(t)\delta_{D}(\boldsymbol{r}-\boldsymbol{r}(t))\rangle$
in Eq.(\ref{eq:evolution_equation}) using Eq. (\ref{eq:Novikov}):
\begin{eqnarray}
\langle\eta_{i}^{\alpha}(t)\delta_{D}(\boldsymbol{r}-\boldsymbol{r}(t))\rangle & = & \int_{-\infty}^{t}\text{d}s\langle\eta_{i}^{\alpha}(t)\eta_{j}^{\beta}(s)\rangle\langle\frac{\delta[\delta_{D}(\boldsymbol{r}-\boldsymbol{r}(t))]}{\delta\eta_{j}^{\beta}(s)}\rangle\nonumber \\
 & = & \frac{D_{a}}{\tau_{p}}\int_{-\infty}^{t}\text{d}s\delta^{\alpha\beta}\delta_{ij}e^{-|t-s|/\tau_{p}}\langle\frac{\delta[\delta_{D}(\boldsymbol{r}-\boldsymbol{r}(t))]}{\delta\eta_{j}^{\beta}(s)}\rangle\nonumber \\
 & = & \frac{D_{a}}{\tau_{p}}\int_{-\infty}^{t}\text{d}s\text{e}^{-|t-s|/\tau_{p}}\langle\frac{\delta[\delta_{D}(\boldsymbol{r}-\boldsymbol{r}(t))]}{\delta\eta_{i}^{\alpha}(s)}\rangle\nonumber \\
 & = & \frac{D_{a}}{\tau_{p}}\int_{-\infty}^{t}\text{d}s\text{e}^{-|t-s|/\tau_{p}}\langle\frac{\partial[\delta_{D}(\boldsymbol{r}-\boldsymbol{r}(t))]}{\partial r_{j}^{\beta}(t)}\frac{\delta r_{j}^{\beta}(t)}{\delta\eta_{i}^{\alpha}(s)}\rangle\nonumber \\
 & = & -\frac{D_{a}}{\tau_{p}}\frac{\partial}{\partial r_{j}^{\beta}}\int_{-\infty}^{t}\text{d}s\text{e}^{-|t-s|/\tau_{p}}\langle\delta_{D}(\boldsymbol{r}-\boldsymbol{r}(t))\frac{\delta r_{j}^{\beta}(t)}{\delta\eta_{i}^{\alpha}(s)}\rangle.\label{eq:term_3_immediate}
\end{eqnarray}
In the second line, we have used the correlation identity $\langle\eta_{i}^{\alpha}(t)\eta_{j}^{\beta}(s)\rangle=\frac{D_{a}}{\tau_{p}}\delta^{\alpha\beta}\delta_{ij}\text{e}^{-|t-s|/\tau_{p}}$.
We introduce two tensors defined by 
\begin{equation}
A_{ji}^{\beta\alpha}(t,s)=\frac{\delta r_{j}^{\beta}(t)}{\delta\eta_{i}^{\alpha}(s)},\;\;\;\;B_{jk}^{\beta\gamma}(t)=\frac{\partial F_{j}^{\beta}}{\partial r_{k}^{\gamma}}\Bigg|_{\boldsymbol{r}(t)}.
\end{equation}
The time derivative of $\boldsymbol{A}$ satisfies
\begin{eqnarray}
\frac{\text{d}A_{ji}^{\beta\alpha}(t,s)}{\text{d}t} & = & \frac{\delta\dot{r}_{j}^{\beta}(t)}{\delta\eta_{i}^{\alpha}(s)}=\frac{\delta}{\delta\eta_{i}^{\alpha}(s)}\left[F_{j}^{\beta}(\boldsymbol{r}(t))+\xi_{j}^{\beta}(t)+\eta_{j}^{\beta}(t)\right]\nonumber \\
 & = & \frac{\partial F_{j}^{\beta}}{\partial r_{k}^{\gamma}}\Bigg|_{\boldsymbol{r}(t)}\frac{\delta r_{k}^{\gamma}(t)}{\delta\eta_{i}^{\alpha}(s)}+\delta^{\alpha\beta}\delta_{ij}\delta_{D}(t-s)\nonumber \\
 & = & B_{jk}^{\beta\gamma}(t)A_{ki}^{\gamma\alpha}(t,s)+\delta^{\alpha\beta}\delta_{ij}\delta_{D}(t-s).\label{eq:A_equation}
\end{eqnarray}
This equation constitutes an ordinary differential equation governing
the tensor $\boldsymbol{A}.$ For $t>s$, the solution can be written
as 
\begin{eqnarray}
A_{ji}^{\beta\alpha}(t,s) & = & \frac{\delta r_{j}^{\beta}(t)}{\delta\eta_{i}^{\alpha}(s)}=\left[\exp\left(\int_{s}^{t}\boldsymbol{B}(s)ds\right)\right]_{ji}^{\beta\alpha}\nonumber \\
 & \approx & \left(e^{(t-s)\boldsymbol{B}(t)}\right)_{ji}^{\beta\alpha}.\label{eq:color_variation}
\end{eqnarray}
In the last step, we have assumed that $\boldsymbol{B}(t)$ varies
slowly on the timescale of the noise correlation time $\tau_{p}$,
which constitutes the central assumption of the Fox approximation
\citep{FOX,FOX_multidimensional_diagnal,FOX_multidimensional_offdiagnal}.
Substituting Eq. (\ref{eq:color_variation}) into Eq. (\ref{eq:term_3_immediate})
yields

\begin{eqnarray}
\langle\eta_{i}^{\alpha}(t)\delta_{D}(\boldsymbol{r}-\boldsymbol{r}(t))\rangle & = & -\frac{D_{a}}{\tau_{p}}\frac{\partial}{\partial r_{j}^{\beta}}\int_{-\infty}^{t}\text{d}se^{-(t-s)/\tau_{p}}\left(e^{(t-s)B}\right)_{ji}^{\beta\alpha}\rho(\boldsymbol{r},t)\nonumber \\
 & = & -\frac{D_{a}}{\tau_{p}}\frac{\partial}{\partial r_{j}^{\beta}}\rho(\boldsymbol{r},t)\left[\int_{0}^{\infty}\text{d}se^{-s(\frac{\mathbf{I}}{\tau_{p}}-\boldsymbol{B})}\right]_{ji}^{\beta\alpha}\nonumber \\
 & = & -\frac{D_{a}}{\tau_{p}}\frac{\partial}{\partial r_{j}^{\beta}}\rho(\boldsymbol{r},t)\left[(\frac{\mathbf{I}}{\tau_{p}}-\boldsymbol{B})^{-1}\right]_{ji}^{\beta\alpha}\nonumber \\
 & = & -D_{a}\frac{\partial}{\partial r_{j}^{\beta}}\rho(\boldsymbol{r},t)\left[(\mathbf{I}-\tau_{p}\boldsymbol{B})^{-1}\right]_{ji}^{\beta\alpha}\nonumber \\
 & = & -D_{a}\nabla_{j}^{\beta}\left[(\mathbf{I}+\tau_{p}\nabla\nabla U)^{-1}\right]_{ij}^{\alpha\beta}\rho(\boldsymbol{r},t).\label{eq:term_3}
\end{eqnarray}
In the final line, we have assumed that the force $\boldsymbol{F}$
is conservative, i.e., $\boldsymbol{F}=-\nabla U$.

Finally, substituting Eqs. (\ref{eq:term_2}) and (\ref{eq:term_3})
into Eq. (\ref{eq:evolution_equation}) yields the probability distribution
evolution equation: 

\begin{equation}
\frac{\partial\rho(\boldsymbol{r},t)}{\partial t}=-\nabla_{i}\cdot\boldsymbol{J}_{i},\label{eq:evoluation_equation}
\end{equation}
where

\begin{eqnarray}
J_{i}^{\alpha} & = & -\left(\nabla_{i}^{\alpha}U\right)\rho(\boldsymbol{r},t)-\nabla_{i}^{\alpha}\rho(\boldsymbol{r},t)-D_{a}\nabla_{j}^{\beta}\left[(\mathbf{I}+\tau_{p}\nabla\nabla U)^{-1}\right]_{ij}^{\alpha\beta}\rho(\boldsymbol{r},t).
\end{eqnarray}

\section{Dissipative work in geometric form}

\subsection{Mean input work}

We begin with the standard definition of the mean input work in stochastic
thermodynamics \citep{jarzynski1997nonequilibrium,sekimoto1997complementarity}
: 
\begin{equation}
W\equiv\int_{0}^{\tau}\text{d}t\langle\frac{\partial U}{\partial t}\rangle=\Delta\langle U\rangle-\int_{0}^{\tau}\text{d}t\langle\frac{\partial U(\boldsymbol{r}(t))}{\partial r_{i}^{\alpha}(t)}\dot{r}_{i}^{\alpha}(t)\rangle.\label{eq:mean_input_work_1}
\end{equation}
Substituting Eq. (\ref{eq:SMAOUP}) into the term $\langle\frac{\partial U(\boldsymbol{r}(t))}{\partial r_{i}^{\alpha}(t)}\dot{r}_{i}^{\alpha}(t)\rangle$
yields 
\begin{eqnarray}
\langle\frac{\partial U(\boldsymbol{r}(t))}{\partial r_{i}^{\alpha}(t)}\dot{r}_{i}^{\alpha}(t)\rangle & = & -\langle\left(\frac{\partial U(\boldsymbol{r}(t))}{\partial r_{i}^{\alpha}(t)}\right)^{2}\rangle+\langle\frac{\partial U(\boldsymbol{r}(t))}{\partial r_{i}^{\alpha}(t)}\xi_{i}^{\alpha}(t)\rangle+\langle\frac{\partial U(\boldsymbol{r}(t))}{\partial r_{i}^{\alpha}(t)}\eta_{i}^{\alpha}(t)\rangle.\label{eq:work_related}
\end{eqnarray}

We again apply Eq. (\ref{eq:Novikov}) to evaluate the second and
third terms in Eq. (\ref{eq:work_related}). First, we consider the
term $\langle\frac{\partial U(\boldsymbol{r}(t))}{\partial r_{i}^{\alpha}(t)}\xi_{i}^{\alpha}(t)\rangle$:
\begin{eqnarray}
\langle\frac{\partial U(\boldsymbol{r}(t))}{\partial r_{i}^{\alpha}(t)}\xi_{i}^{\alpha}(t)\rangle & = & \int_{-\infty}^{t}\text{d}s\langle\xi_{i}^{\alpha}(t)\xi_{j}^{\beta}(s)\rangle\langle\frac{\delta}{\delta\xi_{j}^{\beta}(s)}\frac{\partial U(\boldsymbol{r}(t))}{\partial r_{i}^{\alpha}(t)}\rangle\nonumber \\
 & = & 2\int_{-\infty}^{t}\text{d}s\delta^{\alpha\beta}\delta_{ij}\delta_{D}(t-s)\langle\frac{\partial^{2}U(\boldsymbol{r}(t))}{\partial r_{k}^{\gamma}(t)\partial r_{i}^{\alpha}(t)}\frac{\delta r_{k}^{\gamma}(t)}{\delta\xi_{j}^{\beta}(s)}\rangle\nonumber \\
 & = & 2\int_{-\infty}^{t}\text{d}s\delta^{\alpha\beta}\delta_{ij}\delta_{D}(t-s)\langle\frac{\partial^{2}U(\boldsymbol{r}(t))}{\partial r_{k}^{\gamma}(t)\partial r_{i}^{\alpha}(t)}\theta(t-s)\delta^{\beta\gamma}\delta_{jk}\rangle\nonumber \\
 & = & 2\int_{-\infty}^{t}\text{d}s\delta_{D}(t-s)\langle\frac{\partial^{2}U(\boldsymbol{r}(t))}{\partial r_{i}^{\alpha}(t)^{2}}\theta(t-s)\rangle\nonumber \\
 & = & \langle\frac{\partial^{2}U(\boldsymbol{r}(t))}{\partial r_{i}^{\alpha}(t)^{2}}\rangle.\label{eq:work_related_1}
\end{eqnarray}
Here, we have used $\langle\xi_{i}^{\alpha}(t)\xi_{j}^{\beta}(s)\rangle=2\delta^{\alpha\beta}\delta_{ij}\delta(t-s)$
in the second line and Eq. (\ref{eq:white_variation}) in the third
line. Next, we evaluate the term $\langle\frac{\partial U(\boldsymbol{r}(t))}{\partial r_{i}^{\alpha}(t)}\eta_{i}^{\alpha}(t)\rangle$:
\begin{eqnarray}
\langle\frac{\partial U(\boldsymbol{r}(t))}{\partial r_{i}^{\alpha}(t)}\eta_{i}^{\alpha}(t)\rangle & = & \int_{-\infty}^{t}\text{d}s\langle\eta_{i}^{\alpha}(t)\eta_{j}^{\beta}(s)\rangle\langle\frac{\delta}{\delta\eta_{j}^{\beta}(s)}\frac{\partial U(\boldsymbol{r}(t))}{\partial r_{i}^{\alpha}(t)}\rangle\nonumber \\
 & = & \frac{D_{a}}{\tau_{p}}\int_{-\infty}^{t}\text{d}s\delta^{\alpha\beta}\delta_{ij}e^{-|t-s|/\tau_{p}}\langle\frac{\partial^{2}U(\boldsymbol{r}(t))}{\partial r_{i}^{\alpha}(t)\partial r_{k}^{\gamma}(t)}\frac{\delta r_{k}^{\gamma}(t)}{\delta\eta_{j}^{\beta}(s)}\rangle\nonumber \\
 & = & \frac{D_{a}}{\tau_{p}}\int_{-\infty}^{t}\text{d}s\delta^{\alpha\beta}\delta_{ij}e^{-|t-s|/\tau_{p}}\langle\frac{\partial^{2}U(\boldsymbol{r}(t))}{\partial r_{i}^{\alpha}(t)\partial r_{k}^{\gamma}(t)}\left(e^{(t-s)\boldsymbol{B}}\right)_{kj}^{\gamma\beta}\rangle\nonumber \\
 & = & \frac{D_{a}}{\tau_{p}}\int_{-\infty}^{t}\text{d}se^{-|t-s|/\tau_{p}}\langle\frac{\partial^{2}U(\boldsymbol{r}(t))}{\partial r_{i}^{\alpha}(t)\partial r_{k}^{\gamma}(t)}\left(e^{(t-s)\boldsymbol{B}}\right)_{ki}^{\gamma\alpha}\rangle\nonumber \\
 & = & \frac{D_{a}}{\tau_{p}}\langle\frac{\partial^{2}U(\boldsymbol{r}(t))}{\partial r_{i}^{\alpha}(t)\partial r_{k}^{\gamma}(t)}\left[\int_{0}^{\infty}\text{d}se^{-s(\mathbf{I}/\tau_{p}-\boldsymbol{B})}\right]_{ki}^{\gamma\alpha}\rangle\nonumber \\
 & = & D_{a}\langle\frac{\partial^{2}U(\boldsymbol{r}(t))}{\partial r_{i}^{\alpha}(t)\partial r_{k}^{\gamma}(t)}\left[\left(\mathbf{I}+\tau_{p}\nabla\nabla U\right)^{-1}\right]_{ki}^{\gamma\alpha}\rangle.\label{eq:work_related_2}
\end{eqnarray}
Here, we have used $\langle\eta_{i}^{\alpha}(t)\eta_{j}^{\beta}(t')\rangle=(D_{a}/\tau_{p})\delta^{\alpha\beta}\delta_{ij}\mathrm{e}^{-|t-t'|/\tau_{p}}$
in the second line and Eq. (\ref{eq:color_variation}) in the third.
Substituting Eqs. (\ref{eq:work_related_1}) and (\ref{eq:work_related_2})
into Eq. (\ref{eq:work_related}), we obtain:
\begin{eqnarray}
\langle\frac{\partial U(\boldsymbol{r}(t))}{\partial r_{i}^{\alpha}(t)}\dot{r_{i}^{\alpha}}(t)\rangle & = & -\langle\left(\frac{\partial U(\boldsymbol{r}(t))}{\partial r_{i}^{\alpha}(t)}\right)^{2}\rangle+\langle\frac{\partial^{2}U(\boldsymbol{r}(t))}{\partial r_{i}^{\alpha}(t)^{2}}\rangle+D_{a}\langle\frac{\partial^{2}U(\boldsymbol{r}(t))}{\partial r_{i}^{\alpha}(t)\partial r_{k}^{\gamma}(t)}\left[\left(\mathbf{I}+\tau_{p}\nabla\nabla U\right)^{-1}\right]_{ki}^{\gamma\alpha}\rangle\nonumber \\
 & = & -\langle\left(\frac{\partial U(\boldsymbol{r}(t))}{\partial r_{i}^{\alpha}(t)}\right)^{2}\rangle+\int\text{d}\boldsymbol{r}\left[\frac{\partial}{\partial r_{i}^{\alpha}}\left(\frac{\partial U}{\partial r_{i}^{\alpha}}\rho(\boldsymbol{r},t)\right)-\frac{\partial U}{\partial r_{i}^{\alpha}}\frac{\partial\rho(\boldsymbol{r},t)}{\partial r_{i}^{\alpha}}\right]\nonumber \\
 & + & D_{a}\int\text{d}\boldsymbol{r}\left\{ \frac{\partial}{\partial r_{i}^{\alpha}}\left[\frac{\partial U}{\partial r_{k}^{\gamma}}\left[\left(\mathbf{I}+\tau_{p}\nabla\nabla U\right)^{-1}\right]_{ki}^{\gamma\alpha}\rho(\boldsymbol{r},t)\right]-\frac{\partial U}{\partial r_{k}^{\gamma}}\frac{\partial}{\partial r_{i}^{\alpha}}\left[\left[\left(\mathbf{I}+\tau_{p}\nabla\nabla U\right)^{-1}\right]_{ki}^{\gamma\alpha}\rho(\boldsymbol{r},t)\right]\right\} \nonumber \\
 & = & -\langle\left(\frac{\partial U}{\partial r_{i}^{\alpha}}\right)^{2}\rangle-\int\text{d}\boldsymbol{r}\frac{\partial U}{\partial r_{i}^{\alpha}}\left\{ \frac{\partial\rho}{\partial r_{i}^{\alpha}}+D_{a}\frac{\partial}{\partial r_{k}^{\gamma}}\left[\left[\left(\mathbf{I}+\tau_{p}\nabla\nabla U\right)^{-1}\right]_{ik}^{\alpha\gamma}\rho(\boldsymbol{r},t)\right]\right\} \nonumber \\
 & = & -\langle\left(\frac{\partial U}{\partial r_{i}^{\alpha}}\right)^{2}\rangle+\int\text{d}\boldsymbol{r}\left[\frac{\partial U}{\partial r_{i}^{\alpha}}J_{i}^{\alpha}+\left(\frac{\partial U}{\partial r_{i}^{\alpha}}\right)^{2}\right]\nonumber \\
 & = & \int\text{d}\boldsymbol{r}\frac{\partial U}{\partial r_{i}^{\alpha}}J_{i}^{\alpha}\nonumber \\
 & = & -\int\text{d}\boldsymbol{r}\frac{\boldsymbol{J}_{i}^{2}}{\rho}-\int\text{d}\boldsymbol{r}\frac{J_{i}^{\alpha}}{\rho}\left\{ \frac{\partial\rho}{\partial r_{i}^{\alpha}}+D_{a}\frac{\partial}{\partial r_{j}^{\beta}}\left[\left[\left(I+\tau_{p}\nabla\nabla U\right)^{-1}\right]_{ij}^{\alpha\beta}\rho\right]\right\} .\label{eq:work_related_final}
\end{eqnarray}

Finally, inserting Eq. (\ref{eq:work_related_final}) in Eq. (\ref{eq:mean_input_work_1})
and assuming a short persistence time $\tau_{p}$, the mean input
work can be rewritten as:
\begin{equation}
W=\Delta\langle U\rangle-(1+D_{a})\Delta S+\int_{0}^{\tau}\text{d}t\int d\boldsymbol{r}\frac{\boldsymbol{J}_{i}^{2}}{\rho}-D_{a}\tau_{p}\int_{0}^{\tau}\text{d}t\int\text{d}\boldsymbol{r}\frac{J_{i}^{\alpha}}{\rho}\frac{\partial}{\partial r_{j}^{\beta}}\left(\frac{\partial^{2}U}{\partial r_{i}^{\alpha}\partial r_{j}^{\beta}}\rho\right)+\mathcal{O}(\tau_{p}^{3}),\label{eq:mean_input_work_final}
\end{equation}
where we have used the time derivative of the Gibbs entropy $S(t)=-\int d\boldsymbol{r}\rho\log\rho$:
\begin{eqnarray}
\dot{S}(t) & = & -\int\text{d}\boldsymbol{r}\frac{\partial\rho}{\partial t}\log\rho\nonumber \\
 & = & \int\text{d}\boldsymbol{r}\frac{\partial J_{i}^{\alpha}}{\partial r_{i}^{\alpha}}\log\rho\nonumber \\
 & = & -\int\text{d}\boldsymbol{r}\frac{J_{i}^{\alpha}}{\rho}\frac{\partial\rho}{\partial r_{i}^{\alpha}}.
\end{eqnarray}
Equation (\ref{eq:mean_input_work_final}) expresses the mean input
work as the sum of three terms. The first two terms depend only on
the boundary states, while the remaining terms define the dissipative
work.

\subsection{First-order geometric structure}

We now consider the task of achieving a state transition from an initial
distribution $\rho_{i}$ to a final distribution $\rho_{f}$. To this
end, we design an evolutionary path connecting these states, defined
as:
\begin{equation}
\rho_{B}(\boldsymbol{r},\boldsymbol{\lambda}(t))\equiv\exp\{\frac{F(\boldsymbol{\lambda}(t))-U_{o}(\boldsymbol{r},\boldsymbol{\lambda}(t))}{1+D_{a}}\},\label{eq:path}
\end{equation}
with $\rho_{i}=\rho_{B}(\boldsymbol{r},\boldsymbol{\lambda}(0))$
and $\rho_{f}=\rho_{B}(\boldsymbol{r},\boldsymbol{\lambda}(\tau))$.
Here, $U_{o}(\boldsymbol{r},\boldsymbol{\lambda}(t))\equiv-(1+D_{a})\log\rho_{B}(\boldsymbol{r},\boldsymbol{\lambda}(t))$
is termed the original potential. The vector $\boldsymbol{\lambda}(t)$
represents a set of tunable parameters, $\tau$ denotes the total
driving time, and $F$ is a normalization factor. To guide the system
evolution along the predefined path, an auxiliary potential $U_{a}$
must be introduced which relies on inversely solving the probability
distribution evolution equation.

Under the first-order approximation (i.e. keeping terms up to first
order in $\tau_{p}$), the probability distribution evolution equation
in Eq. (\ref{eq:evoluation_equation}) simplifies to 
\begin{equation}
\frac{\partial\rho}{\partial t}=\nabla_{i}\cdot[(\nabla_{i}U)\rho+(1+D_{a})\nabla_{i}\rho].
\end{equation}
Substituting the total potential $U\equiv U_{o}+U_{a}^{(1)}$ and
the prescribed path $\rho_{B}$ into the equation above yields the
equation that determines the first-order auxiliary potential $U_{a}^{(1)}$:
\begin{equation}
(1+D_{a})\nabla^{2}U_{a}^{(1)}-\nabla_{i}U_{a}^{(1)}\cdot\nabla_{i}U_{o}=\left(\frac{\partial F}{\partial\lambda_{\mu}}-\frac{\partial U_{o}}{\partial\lambda_{\mu}}\right)\dot{\lambda}_{\mu}.\label{eq:evoluation_equation_first}
\end{equation}
By comparing both sides, we obtain the general form of $U_{a}^{(1)}$:
\begin{equation}
U_{a}^{(1)}(\boldsymbol{r},\boldsymbol{\lambda},\dot{\boldsymbol{\lambda}})=\dot{\boldsymbol{\lambda}}\cdot\boldsymbol{f}(\boldsymbol{r},\boldsymbol{\lambda}).\label{eq:U_a_first_form}
\end{equation}

Based on Eq. (\ref{eq:mean_input_work_1}), the mean input work in
such a process is 
\begin{eqnarray}
W^{(1)} & = & \Delta\langle U\rangle-(1+D_{a})\Delta S+\int_{0}^{\tau}\text{d}t\int\text{d}\boldsymbol{r}\frac{(J_{i}^{\alpha})^{2}}{\rho_{B}}.\label{eq:work_1}
\end{eqnarray}
 The probability current under the first-order approximation is given
by
\begin{align}
J_{i}^{\alpha} & =-\left[\frac{\partial U}{\partial r_{i}^{\alpha}}\rho_{B}+(1+D_{a})\frac{\partial\rho_{B}}{\partial r_{i}^{\alpha}}\right]\nonumber \\
 & =-\frac{\partial(U_{o}+U_{a}^{(1)})}{\partial r_{i}^{\alpha}}\rho_{B}+\frac{\partial U_{o}}{\partial r_{i}^{\alpha}}\rho_{B}\nonumber \\
 & =-\frac{\partial U_{a}^{(1)}}{\partial r_{i}^{\alpha}}\rho_{B}.\label{eq:current_first}
\end{align}
Substituting Eqs. (\ref{eq:U_a_first_form}) and ($\ref{eq:current_first}$)
into Eq. (\ref{eq:work_1}), we obtain 
\begin{eqnarray*}
W^{(1)} & = & \Delta\langle U\rangle-(1+D_{a})\Delta S+\int_{0}^{\tau}\text{d}t\int\text{d}\boldsymbol{r}\frac{(J_{i}^{\alpha})^{2}}{\rho_{B}}\\
 & = & \Delta\langle U\rangle-(1+D_{a})\Delta S+\int_{0}^{\tau}\text{d}t\int\text{d}\boldsymbol{r}\rho_{B}\left(\frac{\partial U_{a}^{(1)}}{\partial r_{i}^{\alpha}}\right)^{2}\\
 & = & \Delta\langle U\rangle-(1+D_{a})\Delta S+\int_{0}^{\tau}\text{d}t\int\text{d}\boldsymbol{r}\dot{\lambda}_{\mu}\frac{\partial f_{\mu}}{\partial r_{i}^{\alpha}}\dot{\lambda}_{\nu}\frac{\partial f_{\nu}}{\partial r_{i}^{\alpha}}\rho_{B}\\
 & = & \Delta\langle U\rangle-(1+D_{a})\Delta S+\int_{0}^{\tau}\text{d}t\dot{\lambda}_{\mu}\dot{\lambda}_{\nu}g_{\mu\nu}^{(1)},
\end{eqnarray*}
where $g_{\mu\nu}^{(1)}\equiv\langle\frac{\partial f_{\mu}}{\partial r_{i}^{\alpha}}\frac{\partial f_{\nu}}{\partial r_{i}^{\alpha}}\rangle$
defines a positive semi-definite metric. Here, the average $\langle...\rangle$
is defined as $\langle A\rangle\equiv\int A(\boldsymbol{r})\rho_{B}(\boldsymbol{r},\boldsymbol{\lambda})\text{d}\boldsymbol{r}$.
So, the dissipative work then reads in a geometric form:
\begin{equation}
W_{\text{d}}^{(1)}=\int_{0}^{\tau}\text{d}t\dot{\lambda}_{\mu}\dot{\lambda}_{\nu}g_{\mu\nu}^{(1)}
\end{equation}
with the metric
\begin{equation}
g_{\mu\nu}^{(1)}=\langle\left(\nabla_{i}f_{\mu}\right)\cdot\left(\nabla_{i}f_{\nu}\right)\rangle.
\end{equation}

\subsection{Second-order geometric structure}

Inserting $U\equiv U_{o}+U_{a}^{(2)}$ and Eq. (\ref{eq:path}) into
Eq. (\ref{eq:evoluation_equation}) and retaining terms up to second
order in $\tau_{p}$, we obtain the equation determining the auxiliary
potential $U_{a}^{(2)}$ under the second-order approximation: 
\begin{eqnarray}
\dot{\lambda}_{\mu}\frac{\partial\rho_{B}}{\partial\lambda_{\mu}} & -(1+D_{a})\nabla^{2}\rho_{B}= & \frac{\partial}{\partial r_{i}^{\alpha}}\left(\frac{\partial(U_{o}+U_{a}^{(2)})}{\partial r_{i}^{\alpha}}\rho_{B}\right)-D_{a}\tau_{p}\frac{\partial^{2}}{\partial r_{i}^{\alpha}\partial r_{j}^{\beta}}\left(\frac{\partial^{2}(U_{o}+U_{a}^{(2)})}{\partial r_{i}^{\alpha}\partial r_{j}^{\beta}}\rho_{B}\right).\label{eq:U_a_2}
\end{eqnarray}
By analyzing the structure of Eq. (\ref{eq:U_a_2}), we find $U_{a}^{(2)}$
can be decomposed into two distinct components based on their dependence
on $\dot{\boldsymbol{\lambda}}$: 
\begin{equation}
U_{a}^{(2)}(\boldsymbol{r},\boldsymbol{\lambda},\dot{\boldsymbol{\lambda}})=u(\boldsymbol{r},\boldsymbol{\lambda})+\dot{\boldsymbol{\lambda}}\cdot\boldsymbol{\omega}(\boldsymbol{r},\boldsymbol{\lambda})\label{eq:U_a_second_form}
\end{equation}
Inserting Eq. (\ref{eq:U_a_second_form}) into Eq. (\ref{eq:U_a_2})
and collecting terms without and with $\dot{\boldsymbol{\lambda}}$
yields the equations determining $u$ and $\boldsymbol{\omega}$,
respectively: 
\begin{equation}
D_{a}\tau_{p}\frac{\partial}{\partial r_{j}^{\beta}}\left(\frac{\partial^{2}(U_{o}+u)}{\partial r_{i}^{\alpha}\partial r_{j}^{\beta}}\rho_{B}\right)=\frac{\partial(U_{o}+u)}{\partial r_{i}^{\alpha}}\rho_{B}+(1+D_{a})\frac{\partial\rho_{B}}{\partial r_{i}^{\alpha}},\label{eq:u_equation}
\end{equation}
and
\begin{equation}
\frac{\partial\rho_{B}}{\partial\lambda_{\mu}}=\frac{\partial}{\partial r_{i}^{\alpha}}\left(\frac{\partial\omega_{\mu}}{\partial r_{i}^{\alpha}}\rho_{B}\right)-D_{a}\tau_{p}\frac{\partial^{2}}{\partial r_{i}^{\alpha}\partial r_{j}^{\beta}}\left(\frac{\partial^{2}\omega_{\mu}}{\partial r_{i}^{\alpha}\partial r_{j}^{\beta}}\rho_{B}\right).\label{eq:omega_equation}
\end{equation}
From the Eq. (\ref{eq:omega_equation}), we can express the probability
current in such form: 
\begin{equation}
J_{i}^{\alpha}=\dot{\lambda}_{\mu}h_{i\mu}^{\alpha}\label{eq:current_second}
\end{equation}
with
\begin{equation}
h_{i\mu}^{\alpha}=-\left[\frac{\partial\omega_{\mu}}{\partial r_{i}^{\alpha}}\rho_{B}+D_{a}\tau_{p}\frac{\partial}{\partial r_{j}^{\beta}}\left(\frac{\partial^{2}\omega_{\mu}}{\partial r_{i}^{\alpha}\partial r_{j}^{\beta}}\rho_{B}\right)\right].
\end{equation}

Based on Eq. (\ref{eq:mean_input_work_1}), the mean input work in
such a process can be written as 
\begin{eqnarray}
W^{(2)} & = & \Delta\langle U\rangle-(1+D_{a})\Delta S+\int_{0}^{\tau}\text{d}t\int\text{d}\boldsymbol{r}\frac{(J_{i}^{\alpha})^{2}}{\rho_{B}}\nonumber \\
 &  & -\int_{0}^{\tau}\text{d}t\int\text{d}\boldsymbol{r}\frac{J_{i}^{\alpha}}{\rho_{B}}\frac{\partial}{\partial r_{j}^{\beta}}\left(\frac{\partial^{2}(U_{o}+U_{a}^{(2)})}{\partial r_{i}^{\alpha}\partial r_{j}^{\beta}}\rho_{B}\right).\label{eq:work_2}
\end{eqnarray}
Inserting Eqs. (\ref{eq:U_a_second_form}) and (\ref{eq:current_second})
into Eq. (\ref{eq:work_2}) gives
\begin{eqnarray}
W^{(2)} & = & \Delta\langle U\rangle-(1+D_{a})\Delta S+\int_{0}^{\tau}\text{d}t\int\text{d}\boldsymbol{r}\frac{(J_{i}^{\alpha})^{2}}{\rho_{B}}-D_{a}\tau_{p}\int_{0}^{\tau}\text{d}t\int\text{d}\boldsymbol{r}\frac{J_{i}^{\alpha}}{\rho}\frac{\partial}{\partial r_{j}^{\beta}}\left(\frac{\partial^{2}(U_{o}+U_{a}^{(2)})}{\partial r_{i}^{\alpha}\partial r_{j}^{\beta}}\rho_{B}\right)\nonumber \\
 & = & \Delta\langle U\rangle-(1+D_{a})\Delta S-D_{a}\tau_{p}\int_{0}^{\tau}\text{d}t\dot{\lambda}_{\mu}\int\text{d}\boldsymbol{r}\frac{h_{i\mu}^{\alpha}}{\rho}\frac{\partial}{\partial r_{j}^{\beta}}\left(\frac{\partial^{2}(U_{o}+u)}{\partial r_{i}^{\alpha}\partial r_{j}^{\beta}}\rho_{B}\right)\nonumber \\
 &  & -D_{a}\tau_{p}\int_{0}^{\tau}\text{d}t\dot{\lambda}_{\mu}\dot{\lambda}_{\nu}\int\text{d}\boldsymbol{r}\frac{h_{i\mu}^{\alpha}}{\rho}\frac{\partial}{\partial r_{j}^{\beta}}\left(\frac{\partial^{2}\omega_{\nu}}{\partial r_{i}^{\alpha}\partial r_{j}^{\beta}}\rho_{B}\right)+\int_{0}^{\tau}\text{d}t\dot{\lambda}_{\mu}\dot{\lambda}_{\nu}\int\text{d}\boldsymbol{r}\frac{h_{i\mu}^{\alpha}h_{i\nu}^{\alpha}}{\rho_{B}}\nonumber \\
 & = & \Delta\langle U\rangle-(1+D_{a})\Delta S-D_{a}\tau_{p}\int_{0}^{\tau}\text{d}t\dot{\lambda}_{\mu}\int\text{d}\boldsymbol{r}\frac{h_{i\mu}^{\alpha}}{\rho}\frac{\partial}{\partial r_{j}^{\beta}}\left(\frac{\partial^{2}(U_{o}+u)}{\partial r_{i}^{\alpha}\partial r_{j}^{\beta}}\rho_{B}\right)+\int_{0}^{\tau}\text{d}t\dot{\lambda}_{\mu}\dot{\lambda}_{\nu}g_{\mu\nu}^{(2)}.\label{eq:work_2_shortcut}
\end{eqnarray}
The explicit form of $g_{\mu\nu}^{(2)}$ is obtained by expanding
$h_{i\mu}^{\alpha}$ and simplifying:
\begin{eqnarray}
g_{\mu\nu}^{(2)} & = & \int\text{d}\boldsymbol{r}\frac{h_{i\mu}^{\alpha}h_{i\nu}^{\alpha}}{\rho_{B}}-D_{a}\tau_{p}\int\text{d}\boldsymbol{r}\frac{h_{i\mu}^{\alpha}}{\rho}\frac{\partial}{\partial r_{j}^{\beta}}\left(\frac{\partial^{2}\omega_{\nu}}{\partial r_{i}^{\alpha}\partial r_{j}^{\beta}}\rho_{B}\right)\nonumber \\
 & = & \int\text{d}\boldsymbol{r}\frac{h_{i\mu}^{\alpha}h_{i\nu}^{\alpha}}{\rho_{B}}-D_{a}\tau_{p}\int\text{d}\boldsymbol{r}\frac{\partial\omega_{\mu}}{\partial r_{i}^{\alpha}}\frac{\partial}{\partial r_{j}^{\beta}}\left(\frac{\partial^{2}\omega_{\nu}}{\partial r_{i}^{\alpha}\partial r_{j}^{\beta}}\rho_{B}\right)\nonumber \\
 &  & -D_{a}^{2}\tau_{p}^{2}\int\text{d}\boldsymbol{r}\frac{\partial}{\partial r_{j}^{\beta}}\left(\frac{\partial^{2}\omega_{\mu}}{\partial r_{i}^{\alpha}\partial r_{j}^{\beta}}\rho_{B}\right)\frac{\partial}{\partial r_{k}^{\gamma}}\left(\frac{\partial^{2}\omega_{\nu}}{\partial r_{i}^{\alpha}\partial r_{k}^{\gamma}}\rho_{B}\right)\nonumber \\
 & = & \int\text{d}\boldsymbol{r}\frac{\partial\omega_{\mu}}{\partial r_{i}^{\alpha}}\frac{\partial\omega_{\nu}}{\partial r_{i}^{\alpha}}\rho_{B}+D_{a}\tau_{p}\int\text{d}\boldsymbol{r}\frac{\partial\omega_{\mu}}{\partial r_{i}^{\alpha}}\frac{\partial}{\partial r_{j}^{\beta}}\left(\frac{\partial^{2}\omega_{\nu}}{\partial r_{i}^{\alpha}\partial r_{j}^{\beta}}\rho_{B}\right)\nonumber \\
 &  & +D_{a}\tau_{p}\int\text{d}\boldsymbol{r}\frac{\partial\omega_{\nu}}{\partial r_{i}^{\alpha}}\frac{\partial}{\partial r_{j}^{\beta}}\left(\frac{\partial^{2}\omega_{\mu}}{\partial r_{i}^{\alpha}\partial r_{j}^{\beta}}\rho_{B}\right)-D_{a}\tau_{p}\int\text{d}\boldsymbol{r}\frac{\partial\omega_{\mu}}{\partial r_{i}^{\alpha}}\frac{\partial}{\partial r_{j}^{\beta}}\left(\frac{\partial^{2}\omega_{\nu}}{\partial r_{i}^{\alpha}\partial r_{j}^{\beta}}\rho_{B}\right)\nonumber \\
 &  & +D_{a}^{2}\tau_{p}^{2}\int\text{d}\boldsymbol{r}\frac{\partial}{\partial r_{j}^{\beta}}\left(\frac{\partial^{2}\omega_{\mu}}{\partial r_{i}^{\alpha}\partial r_{j}^{\beta}}\rho_{B}\right)\frac{\partial}{\partial r_{k}^{\gamma}}\left(\frac{\partial^{2}\omega_{\nu}}{\partial r_{i}^{\alpha}\partial r_{k}^{\gamma}}\rho_{B}\right)\nonumber \\
 &  & -D_{a}^{2}\tau_{p}^{2}\int\text{d}\boldsymbol{r}\frac{\partial}{\partial r_{j}^{\beta}}\left(\frac{\partial^{2}\omega_{\mu}}{\partial r_{i}^{\alpha}\partial r_{j}^{\beta}}\rho_{B}\right)\frac{\partial}{\partial r_{k}^{\gamma}}\left(\frac{\partial^{2}\omega_{\nu}}{\partial r_{i}^{\alpha}\partial r_{k}^{\gamma}}\rho_{B}\right)\nonumber \\
 & = & \int\text{d}\boldsymbol{r}\frac{\partial\omega_{\mu}}{\partial r_{i}^{\alpha}}\frac{\partial\omega_{\nu}}{\partial r_{i}^{\alpha}}\rho_{B}+D_{a}\tau_{p}\int\text{d}\boldsymbol{r}\frac{\partial\omega_{\mu}}{\partial r_{i}^{\alpha}}\frac{\partial}{\partial r_{j}^{\beta}}\left(\frac{\partial^{2}\omega_{\nu}}{\partial r_{i}^{\alpha}\partial r_{j}^{\beta}}\rho_{B}\right)\nonumber \\
 & = & \int\text{d}\boldsymbol{r}\frac{\partial\omega_{\mu}}{\partial r_{i}^{\alpha}}\frac{\partial\omega_{\nu}}{\partial r_{i}^{\alpha}}\rho_{B}-D_{a}\tau_{p}\int\text{d}\boldsymbol{r}\left(\frac{\partial^{2}\omega_{\mu}}{\partial r_{i}^{\alpha}\partial r_{j}^{\beta}}\right)\left(\frac{\partial^{2}\omega_{\nu}}{\partial r_{i}^{\alpha}\partial r_{j}^{\beta}}\rho_{B}\right)\nonumber \\
 & = & \langle\frac{\partial\omega_{\mu}}{\partial r_{i}^{\alpha}}\frac{\partial\omega_{\nu}}{\partial r_{i}^{\alpha}}\rangle+D_{a}\tau_{p}\langle\frac{\partial^{2}\omega_{\mu}}{\partial r_{i}^{\alpha}\partial r_{j}^{\beta}}\frac{\partial^{2}\omega_{\nu}}{\partial r_{i}^{\alpha}\partial r_{j}^{\beta}}\rangle.
\end{eqnarray}
In the penultimate line, we have integrated by parts and have assumed
that boundary terms vanish. It follows that $g_{\mu\nu}^{(2)}$ defines
a positive semi-definite metric. Next, we simplify the remaining integral
term in Eq. (\ref{eq:work_2_shortcut}):
\begin{eqnarray}
-D_{a}\tau_{p}\int_{0}^{\tau}\text{d}t\dot{\lambda}_{\mu}\int\text{d}\boldsymbol{r}\frac{h_{i\mu}^{\alpha}}{\rho_{B}}\frac{\partial}{\partial r_{j}^{\beta}}\left(\frac{\partial^{2}(U_{o}+u)}{\partial r_{i}^{\alpha}\partial r_{j}^{\beta}}\rho_{B}\right) & = & -\int_{0}^{\tau}\text{d}t\int\text{d}\boldsymbol{r}\frac{J_{i}^{\alpha}\left[\frac{\partial(U_{o}+u)}{\partial r_{i}^{\alpha}}\rho_{B}+(1+D_{a})\frac{\partial\rho_{B}}{\partial r_{i}^{\alpha}}\right]}{\rho_{B}}\nonumber \\
 & = & -\int_{0}^{\tau}\text{d}t\int\text{d}\boldsymbol{r}\left\{ J_{i}^{\alpha}\frac{\partial(U_{o}+u)}{\partial r_{i}^{\alpha}}+(1+D_{a})J_{i}^{\alpha}\frac{\partial}{\partial r_{i}^{\alpha}}\log\rho_{B}\right\} \nonumber \\
 & = & \int_{0}^{\tau}\text{d}t\int\text{d}\boldsymbol{r}\left\{ \frac{\partial J_{i}^{\alpha}}{\partial r_{i}^{\alpha}}(U_{o}+u)+(1+D_{a})\frac{\partial J_{i}^{\alpha}}{\partial r_{i}^{\alpha}}\log\rho_{B}\right\} \nonumber \\
 & = & -\int_{0}^{\tau}\text{d}t\int\text{d}\boldsymbol{r}\frac{\partial\rho_{B}}{\partial t}\left\{ (U_{o}+u)+(1+D_{a})\log\rho_{B}\right\} \nonumber \\
 & = & -\int_{0}^{\tau}\text{d}t\int\text{d}\boldsymbol{r}\left[\frac{\partial\rho_{B}}{\partial t}(U_{o}+u)\right]+(1+D_{a})\Delta S.
\end{eqnarray}
In the first line we have used Eq. (\ref{eq:u_equation}), and in
the third line we have integrated by parts and assumed vanishing boundary
terms. It is important to note that $U_{o}$ and $u$ can be treated
as functionals of $\rho_{B}$, and the first term in the last line
depends only on the boundary conditions (i.e., $\boldsymbol{\lambda}(0)$
and $\boldsymbol{\lambda}(\tau)$). Therefore, the dissipative work
to second order takes a geometric form:
\begin{equation}
W_{\text{d}}^{(2)}=\int_{0}^{\tau}\text{d}t\dot{\lambda}_{\mu}\dot{\lambda}_{\nu}g_{\mu\nu}^{(2)}
\end{equation}
with the metric
\begin{equation}
g_{\mu\nu}^{(2)}=\langle(\nabla_{i}\omega_{\mu})\cdot(\nabla_{i}\omega_{\nu})\rangle+D_{a}\tau_{p}\langle(\nabla_{i}\nabla_{j}\omega_{\mu})\cdot(\nabla_{i}\nabla_{j}\omega_{\nu})\rangle.
\end{equation}

Compared to the first-order result, the second-order expansion introduces
higher-order spatial derivatives of the auxiliary field, leading to
an additional contribution to the metric. This result reveals that
active fluctuations effectively renormalize the thermodynamic metric
by introducing higher-order spatial correlations, thereby reflecting
the non-Markovian nature of active dynamics.

\section{Applications}

\subsection{Harmonic Coupling}

To illustrate the general framework in an analytically tractable setting,
we first consider an interacting system confined in a harmonic trap,
for which both the auxiliary potential and the thermodynamic metric
can be obtained in closed form. We consider a system confined in a
harmonic trap of strength $\lambda_{1}(t)$, in which each particle
$i$ is coupled to the center of mass $\boldsymbol{R}\equiv\frac{1}{N}\sum_{i}\boldsymbol{r}_{i}$
via the potential 
\begin{equation}
u_{i}=\frac{1}{2}\lambda_{2}(t)(\boldsymbol{r}_{i}-\boldsymbol{R})^{2}.
\end{equation}
 The total potential, including both interactions and the external
confinement, is given by 
\begin{equation}
U_{o}=\frac{1}{2}\lambda_{1}(t)\boldsymbol{r}_{i}^{2}+\frac{1}{2}\lambda_{2}(t)(\boldsymbol{r}_{i}-\boldsymbol{R})^{2}.
\end{equation}
 To facilitate subsequent calculations, we rewrite $U_{o}$ as follows:
\begin{eqnarray}
U_{o} & = & \frac{\lambda_{1}}{2}\boldsymbol{r}_{i}^{2}+\frac{\lambda_{2}}{2}\boldsymbol{r}_{\boldsymbol{i}}^{2}-\frac{\lambda_{2}}{N}\sum_{ij}\boldsymbol{r}_{i}\cdot\boldsymbol{r}_{j}+\frac{\lambda_{2}}{2N^{2}}\sum_{ijk}\boldsymbol{r}_{j}\cdot\boldsymbol{r}_{k}\nonumber \\
 & = & \frac{\lambda_{1}}{2}\boldsymbol{r}_{i}^{2}+\frac{\lambda_{2}}{2}\boldsymbol{r}_{\boldsymbol{i}}^{2}-\frac{\lambda_{2}}{2N}\sum_{ij}\boldsymbol{r}_{i}\cdot\boldsymbol{r}_{j}\nonumber \\
 & = & \frac{\lambda_{1}}{2}\boldsymbol{r}_{i}^{2}+\frac{\lambda_{2}}{2}\left[\left(\boldsymbol{r}^{\alpha}\right)^{\text{T}}\mathbf{I}\boldsymbol{r}^{\alpha}-\frac{1}{N}\sum_{ij}r_{i}^{\alpha}r_{j}^{\alpha}\right]\nonumber \\
 & = & \frac{\lambda_{1}}{2}\left(\boldsymbol{r}^{\alpha}\right)^{\text{T}}\mathbf{I}\boldsymbol{r}^{\alpha}+\frac{\lambda_{2}}{2}\left[\left(\boldsymbol{r}^{\alpha}\right)^{\text{T}}\mathbf{I}\boldsymbol{r}^{\alpha}-\frac{1}{N}r_{i}^{\alpha}\left(\mathbf{11}^{\text{T}}\right)_{ij}r_{j}^{\alpha}\right]\nonumber \\
 & = & \frac{\lambda_{1}}{2}\left(\boldsymbol{r}^{\alpha}\right)^{\text{T}}\mathbf{I}\boldsymbol{r}^{\alpha}+\frac{\lambda_{2}}{2}\left[\left(\boldsymbol{r}^{\alpha}\right)^{\text{T}}\mathbf{I}\boldsymbol{r}^{\alpha}-\frac{1}{N}\left(\boldsymbol{r}^{\alpha}\right)^{\text{T}}\left(\mathbf{11}^{\text{T}}\right)_{ij}\boldsymbol{r}^{\alpha}\right]\nonumber \\
 & = & \frac{\lambda_{1}}{2}\left(\boldsymbol{r}^{\alpha}\right)^{\text{T}}\mathbf{I}\boldsymbol{r}^{\alpha}+\frac{\lambda_{2}}{2}\left(\boldsymbol{r}^{\alpha}\right)^{\text{T}}\mathbf{C}\boldsymbol{r}^{\alpha}\nonumber \\
 & = & \frac{1}{2}\left(\boldsymbol{r}^{\alpha}\right)^{\text{T}}\mathbf{\Lambda}\boldsymbol{r}^{\alpha},\label{eq:U_o_harmonic}
\end{eqnarray}
 where we have defined the column vector $\boldsymbol{r}^{\alpha}\equiv\left[r_{1}^{\alpha},r_{2}^{\alpha},\dots,r_{N}^{\alpha}\right]^{T}$
and introduced the matrix 
\begin{equation}
\mathbf{C}\equiv\mathbf{I}-\frac{1}{N}\mathbf{11}^{\text{T}}.
\end{equation}
 Here, $\mathbf{11}^{\text{T}}$ denotes a matrix with all entries
equal to unity and $\mathbf{I}$ is the identity matrix. Our objective
is to derive explicit expressions for the auxiliary potential, the
thermodynamic metric, and the corresponding geodesic protocols.

We now rewrite the probability distribution evolution equation in
Eq. (\ref{eq:evoluation_equation}) under the first-order approximation
in the following form:
\begin{equation}
\frac{\partial\rho}{\partial t}=\nabla^{\alpha}\cdot\left[\left(\nabla^{\alpha}U\right)\rho+(1+D_{a})\nabla^{\alpha}\rho\right],\label{eq:evoluation_equation_harmonic_first}
\end{equation}
 where $\nabla^{\alpha}\equiv\left[\nabla_{1}^{\alpha},\nabla_{2}^{\alpha},\dots,\nabla_{N}^{\alpha}\right]^{T}$.
Assuming that the system follows the distribution $\rho_{B}(\boldsymbol{r},\mathbf{\Lambda}(t))\equiv\exp\{\frac{F(\mathbf{\Lambda}(t))-U_{o}(\boldsymbol{r}\boldsymbol{,}\mathbf{\Lambda}(t))}{1+D_{a}}\}$,
we obtain 
\begin{equation}
\nabla^{\alpha}\rho_{B}=-\frac{\rho_{B}}{1+D_{a}}\nabla^{\alpha}U_{o}\label{eq:rho_B_first_derivation}
\end{equation}
 and 
\begin{align}
\nabla^{2}\rho_{B} & =\nabla^{\alpha}\cdot\nabla^{\alpha}\rho_{B}=-\frac{\rho_{B}}{1+D_{a}}\nabla^{\alpha}U_{o}+\frac{\rho_{B}}{(1+D_{a})^{2}}\nabla^{\alpha}U_{o}\cdot\nabla^{\alpha}U_{o}\nonumber \\
 & =-\frac{\rho_{B}}{1+D_{a}}\nabla^{\alpha}U_{o}+\frac{\rho_{B}}{(1+D_{a})^{2}}\left(\nabla^{\alpha}U_{o}\right)^{2}.\label{eq:rho_B_second_derivation}
\end{align}
 Substituting $\rho=\rho_{B}$, $U=U_{o}+U_{a}^{(1)}$, Eqs. (\ref{eq:rho_B_first_derivation})
and (\ref{eq:rho_B_second_derivation}) into Eq. (\ref{eq:evoluation_equation_harmonic_first}),
we obtain the equation determining the first-order auxiliary potential
$U_{a}^{(1)}$:
\begin{equation}
\text{Tr}\left[\left(\frac{\partial F}{\partial\mathbf{\Lambda}}-\frac{\partial U_{o}}{\partial\mathbf{\Lambda}}\right)\dot{\mathbf{\Lambda}}\right]=(1+D_{a})\nabla^{2}U_{a}^{(1)}-\nabla^{\alpha}U_{o}\cdot\nabla^{\alpha}U_{a}^{(1)},\label{eq:U_a_first_equation}
\end{equation}
 which corresponds to Eq. (\ref{eq:evoluation_equation_first}). The
normalization factor $F$ is given by
\begin{eqnarray}
F & = & -(1+D_{a})\log\int\exp\left(-\frac{\left(\boldsymbol{r}^{\alpha}\right)^{\text{T}}\mathbf{\Lambda}\boldsymbol{r}^{\alpha}}{2(1+D_{a})}\right)\text{d}\boldsymbol{r}\\
 & = & -(1+D_{a})\log\sqrt{\frac{(2\pi)^{N}(1+D_{a})^{N}}{\det(\mathbf{\Lambda})}}\label{eq:F}
\end{eqnarray}
 and the derivative of $F$ with respect to $\boldsymbol{\Lambda}$
is given by
\begin{equation}
\frac{\partial F}{\partial\mathbf{\Lambda}}=-(1+D_{a})(-\frac{1}{2}\mathbf{\Lambda}^{-1})=\frac{1+D_{a}}{2}\mathbf{\Lambda}^{-1},\label{eq:F_derivative}
\end{equation}
 where we used the identity $\frac{\partial}{\partial\mathbf{\Lambda}}\det(\mathbf{\Lambda})=\det(\mathbf{\Lambda})\mathbf{\Lambda}^{-1}$.
Substituting Eqs. (\ref{eq:U_o_harmonic}), (\ref{eq:F}), and (\ref{eq:F_derivative})
into Eq. (\ref{eq:U_a_first_equation}), we arrive at
\begin{equation}
\nabla^{2}U_{a}^{(1)}-\frac{1}{1+D_{a}}\left(\boldsymbol{r}^{\alpha}\right)^{\text{T}}\mathbf{\Lambda}\nabla^{\alpha}U_{a}^{(1)}=\frac{1}{2}\text{Tr}\left(\mathbf{\Lambda}^{-1}\dot{\mathbf{\Lambda}}\right)-\frac{1}{2(1+D_{a})}\left(\boldsymbol{r}^{\alpha}\right)^{\text{T}}\dot{\mathbf{\Lambda}}\boldsymbol{r}^{\alpha}.
\end{equation}
 The solution of this equation is
\begin{equation}
U_{a}^{(1)}=\frac{1}{4}\left(\boldsymbol{r}^{\alpha}\right)^{\text{T}}\mathbf{\Lambda}^{-1}\dot{\mathbf{\Lambda}}\boldsymbol{r}^{\alpha},
\end{equation}
 so that
\begin{equation}
f_{1}=\frac{1}{4}\left(\boldsymbol{r}^{\alpha}\right)^{\text{T}}\mathbf{\Lambda}^{-1}\boldsymbol{r}^{\alpha},\label{eq:f_1_formula}
\end{equation}
\begin{equation}
f_{2}=\frac{1}{4}\left(\boldsymbol{r}^{\alpha}\right)^{\text{T}}\mathbf{\Lambda}^{-1}\mathbf{C}\boldsymbol{r}^{\alpha}.\label{eq:f_2_formula}
\end{equation}

We now employ a standard identity for quadratic forms of multivariate
Gaussian random variables. For a multivariate Gaussian random vector
\textbf{$\boldsymbol{v}$ }with zero mean and covariance matrix $\mathbf{\Sigma}^{-1}$,
and for any real symmetric matrix $\mathbf{A}$ of compatible dimensions,
the expectation value of the quadratic form $\boldsymbol{v}^{\text{T}}\mathbf{A}\boldsymbol{v}$
is given by 
\begin{equation}
\mathbb{E}(\boldsymbol{v}^{T}\mathbf{A}\boldsymbol{v})=\text{Tr}(\mathbf{A}\mathbf{\Sigma}^{-1}).
\end{equation}
 This identity will be used repeatedly to simplify the calculation
of the metric tensor components 
\begin{equation}
g_{\mu\nu}^{(1)}\equiv\langle\left(\nabla_{i}f_{\mu}\right)\cdot\left(\nabla_{i}f_{\nu}\right)\rangle=\langle\left(\nabla^{\alpha}f_{\mu}\right)\cdot\left(\nabla^{\alpha}f_{\nu}\right)\rangle.
\end{equation}
 We evaluate the  component $g_{22}^{(1)}$ using Eqs. (\ref{eq:f_1_formula})
and (\ref{eq:f_2_formula}) as an illustrative example:

\begin{eqnarray}
g_{22}^{(1)} & = & \langle\left(\nabla^{\alpha}f_{2}\right)\cdot\left(\nabla^{\alpha}f_{2}\right)\rangle=\frac{1}{4}\langle\left(\boldsymbol{r}^{\alpha}\right)^{\text{T}}\mathbf{\Lambda}^{-1}\mathbf{C}\mathbf{\Lambda}^{-1}\boldsymbol{r}^{\alpha}\rangle\nonumber \\
 & = & \frac{1}{4(\lambda_{1}+\lambda_{2})^{2}}\langle\left(\boldsymbol{r}^{\alpha}\right)^{\text{T}}\mathbf{C}\boldsymbol{r}^{\alpha}\rangle\nonumber \\
 & = & \frac{n(1+D_{a})}{4(\lambda_{1}+\lambda_{2})^{2}}\text{Tr}(\mathbf{C}\mathbf{\Lambda}^{-1})\nonumber \\
 & = & \frac{n(1+D_{a})}{4(\lambda_{1}+\lambda_{2})^{3}}\text{Tr}(\mathbf{C})\nonumber \\
 & = & \frac{n(1+D_{a})(N-1)}{4(\lambda_{1}+\lambda_{2})^{3}},
\end{eqnarray}
 where $n$ denotes the spatial dimension and we have used $\mathbf{\Lambda}^{-1}\mathbf{C}=\mathbf{C}\mathbf{\Lambda}^{-1}=\frac{\mathbf{C}}{\lambda_{1}+\lambda_{2}}$
and $\text{Tr}(\mathbf{C})=N-1$, both of which can be verified directly.
By analogous reasoning, the remaining components are found to be 
\begin{equation}
g_{12}^{(1)}=g_{21}^{(1)}=g_{22}^{(1)}
\end{equation}
 and

\begin{equation}
g_{11}^{(1)}=\frac{n(1+D_{a})}{4\lambda_{1}^{3}}+\frac{n(1+D_{a})(N-1)}{4(\lambda_{1}+\lambda_{2})^{3}}.
\end{equation}
 The geodesic equation in its standard form is given by 
\begin{equation}
\frac{d^{2}\lambda_{\mu}}{dt^{2}}+\Gamma_{\nu\sigma}^{\mu}\frac{d\lambda_{\nu}}{dt}\frac{d\lambda_{\sigma}}{dt}=0,
\end{equation}
 where the Christoffel symbols $\Gamma_{\nu\sigma}^{\mu}=\frac{1}{2}(\boldsymbol{g}^{(1)})_{\mu\kappa}^{-1}(\frac{\partial g_{\kappa\sigma}^{(1)}}{\partial\lambda_{\nu}}+\frac{\partial g_{\kappa\nu}^{(1)}}{\partial\lambda_{\sigma}}-\frac{\partial g_{\nu\sigma}^{(1)}}{\partial\lambda_{\kappa}})$.
Substituting the metric components obtained above into the definition
of the Christoffel symbols, and performing straightforward differentiation,
we obtain the following set of differential equations:
\begin{equation}
\ddot{\lambda}_{1}-\frac{3\dot{\lambda}_{1}^{2}}{2\lambda_{1}}=0
\end{equation}
 and 
\begin{equation}
\ddot{\lambda}_{1}+\ddot{\lambda}_{2}-\frac{3(\dot{\lambda}_{1}+\dot{\lambda}_{2})^{2}}{2(\lambda_{1}+\lambda_{2})}=0.
\end{equation}
 These nonlinear ordinary differential equations admit analytical
solutions. The resulting geodesic protocols are given by
\begin{equation}
\lambda_{1}(t)=\left[\lambda_{1i}^{-\frac{1}{2}}+\frac{t}{\tau}\left(\lambda_{1f}^{-\frac{1}{2}}-\lambda_{1i}^{-\frac{1}{2}}\right)\right]^{-2}
\end{equation}
 and 
\begin{equation}
\lambda_{2}(t)=\left\{ \left(\lambda_{1i}+\lambda_{2i}\right)^{-\frac{1}{2}}+\frac{t}{\tau}\left[\left(\lambda_{1f}+\lambda_{2f}\right)^{-\frac{1}{2}}-\left(\lambda_{1i}+\lambda_{2i}\right)^{-\frac{1}{2}}\right]\right\} ^{-2}-\left[\lambda_{1i}^{-\frac{1}{2}}+\frac{t}{\tau}\left(\lambda_{1f}^{-\frac{1}{2}}-\lambda_{1i}^{-\frac{1}{2}}\right)\right]^{-2}.
\end{equation}
 Here, the subscripts $i$ and $f$ denote the initial and final values
of the control parameters, respectively.

We now rewrite the probability distribution evolution equation Eq.
(\ref{eq:evoluation_equation}) under the second-order approximation
as 
\begin{equation}
\frac{\partial\rho}{\partial t}=\nabla^{\alpha}\cdot\left[\left(\nabla^{\alpha}U\right)\rho\right]+(1+D_{a})\nabla^{2}\rho-D_{a}\tau_{p}\nabla^{\alpha}\nabla^{\beta}\cdot\left[\left(\nabla^{\alpha}\nabla^{\beta}U\right)\rho\right].\label{eq:evoluation_equation_harmonic_second}
\end{equation}
 For convenience, we introduce two scalar functions $\tilde{u}=U_{o}+u$
and $\Omega=\dot{\boldsymbol{\lambda}}\cdot\boldsymbol{\omega}$ such
that the total potential can be written as 
\begin{eqnarray}
U & = & U_{o}+U_{a}^{(2)}\nonumber \\
 & = & U_{o}+u+\dot{\boldsymbol{\lambda}}\cdot\boldsymbol{\omega}\nonumber \\
 & = & \tilde{u}+\Omega.\label{eq:U_harmonic_rewritten}
\end{eqnarray}
 Substituting $\rho=\rho_{B}$ and Eq. (\ref{eq:U_harmonic_rewritten})
into Eq. (\ref{eq:evoluation_equation_harmonic_second}), we obtain
the equations determining the second-order auxiliary potential $U_{a}^{(2)}$:
\begin{equation}
\left(\nabla^{\alpha}\tilde{u}\right)\rho_{B}+(1+D_{a})\nabla^{\alpha}\rho_{B}=D_{a}\tau_{p}\nabla^{\beta}\left[\left(\nabla^{\alpha}\nabla^{\beta}\tilde{u}\right)\rho_{B}\right]\label{eq:u_tilde_equation}
\end{equation}
 and 
\begin{equation}
\frac{\partial\rho_{B}}{\partial t}=\nabla^{\alpha}\cdot\left[\left(\nabla^{\alpha}\Omega\right)\rho_{B}\right]-D_{a}\tau_{p}\nabla^{\alpha}\nabla^{\beta}\cdot\left[\left(\nabla^{\alpha}\nabla^{\beta}\Omega\right)\rho_{B}\right].\label{eq:Omega_equation}
\end{equation}
 These equations correspond to Eqs. (\ref{eq:u_equation}) and (\ref{eq:omega_equation}),
respectively. We assume that both $\tilde{u}$ and $\Omega$ take
the quadratic forms: 
\begin{equation}
\tilde{u}=\frac{1}{2}\left(\boldsymbol{r}^{\alpha}\right)^{\text{T}}\mathbf{P}\boldsymbol{r}^{\alpha}\label{eq:u_tilde_form}
\end{equation}
 and 
\begin{equation}
\Omega=\frac{1}{2}\left(\boldsymbol{r}^{\alpha}\right)^{\text{T}}\mathbf{Q}\boldsymbol{r}^{\alpha}.\label{eq:omega_form}
\end{equation}
 Substituting Eqs. (\ref{eq:u_tilde_form}) and (\ref{eq:omega_form})
into Eqs. (\ref{eq:u_tilde_equation}) and (\ref{eq:Omega_equation}),
respectively, and performing straightforward algebraic manipulations,
we obtain 
\begin{equation}
\mathbf{P}=\left(\mathbf{I}+\frac{D_{a}\tau_{p}}{1+D_{a}}\mathbf{\Lambda}\right)^{-1}\mathbf{\Lambda}
\end{equation}
 and 
\begin{equation}
\mathbf{Q}=\frac{1}{2}\left(\mathbf{I}+\frac{D_{a}\tau_{p}}{1+D_{a}}\mathbf{\Lambda}\right)^{-1}\mathbf{\Lambda}^{-1}\dot{\mathbf{\Lambda}}.
\end{equation}
 This structure can be interpreted as a persistence-induced renormalization
of the stiffness matrix. Thus, we arrive at the explicit forms of
$\tilde{u}$ and $\Omega$:
\begin{equation}
\tilde{u}=\frac{1}{2}\left(\boldsymbol{r}^{\alpha}\right)^{\text{T}}\left(\mathbf{I}+\frac{D_{a}\tau_{p}}{1+D_{a}}\mathbf{\Lambda}\right)^{-1}\mathbf{\Lambda}\boldsymbol{r}^{\alpha}
\end{equation}
 and 
\begin{equation}
\Omega=\frac{1}{4}\left(\boldsymbol{r}^{\alpha}\right)^{\text{T}}\left(\mathbf{I}+\frac{D_{a}\tau_{p}}{1+D_{a}}\mathbf{\Lambda}\right)^{-1}\mathbf{\Lambda}^{-1}\dot{\mathbf{\Lambda}}\boldsymbol{r}^{\alpha}.
\end{equation}
 Accordingly, the functions $\omega_{1}$ and $\omega_{2}$ are given
by 
\begin{equation}
\omega_{1}=\frac{1}{4}\left(\boldsymbol{r}^{\alpha}\right)^{\text{T}}\left(\mathbf{I}+\frac{D_{a}\tau_{p}}{1+D_{a}}\mathbf{\Lambda}\right)^{-1}\mathbf{\Lambda}^{-1}\boldsymbol{r}^{\alpha}\label{eq:omega_1_formula}
\end{equation}
 and 
\begin{equation}
\omega_{2}=\frac{1}{4}\left(\boldsymbol{r}^{\alpha}\right)^{\text{T}}\left(\mathbf{I}+\frac{D_{a}\tau_{p}}{1+D_{a}}\mathbf{\Lambda}\right)^{-1}\mathbf{\Lambda}^{-1}\mathbf{C}\boldsymbol{r}^{\alpha}.\label{eq:omega_2_formula}
\end{equation}

As an illustrative example, we evaluate the component $g_{22}$ using
Eqs. (\ref{eq:omega_1_formula}) and (\ref{eq:omega_2_formula}) :
\begin{align}
g_{22}^{(2)} & =\int\text{d}\boldsymbol{r}\left(\nabla^{\alpha}\omega_{2}\right)\cdot\left(\nabla^{\alpha}\omega_{2}\right)\rho_{B}+D_{a}\tau_{p}\int\text{d}\boldsymbol{r}\left(\nabla^{\alpha}\nabla^{\beta}\omega_{2}\right)\cdot\left(\nabla^{\alpha}\nabla^{\beta}\omega_{2}\right)\rho_{B}\nonumber \\
 & =\frac{1}{4}\mathbb{E}\left\{ \left(\boldsymbol{r}^{\alpha}\right)^{\text{T}}\left[\left(\mathbf{I}+\frac{D_{a}\tau_{p}}{1+D_{a}}\mathbf{\Lambda}\right)^{-1}\mathbf{\Lambda}^{-1}\mathbf{C}\right]^{2}\boldsymbol{r}^{\alpha}\right\} \nonumber \\
 & +\frac{nD_{a}\tau_{p}}{4}\text{Tr}\left\{ \left[\left(\mathbf{I}+\frac{D_{a}\tau_{p}}{1+D_{a}}\mathbf{\Lambda}\right)^{-1}\mathbf{\Lambda}^{-1}\mathbf{C}\right]^{2}\right\} \nonumber \\
 & =\frac{n(1+D_{a})}{4}\text{Tr}\left\{ \left[\left(\mathbf{I}+\frac{D_{a}\tau_{p}}{1+D_{a}}\mathbf{\Lambda}\right)^{-1}\mathbf{\Lambda}^{-1}\mathbf{C}\right]^{2}\mathbf{\Lambda}^{-1}\right\} \nonumber \\
 & +\frac{nD_{a}\tau_{p}}{4}\text{Tr}\left\{ \left[\left(\mathbf{I}+\frac{D_{a}\tau_{p}}{1+D_{a}}\mathbf{\Lambda}\right)^{-1}\mathbf{\Lambda}^{-1}\mathbf{C}\right]^{2}\right\} .
\end{align}
 By analogous calculations, the remaining components are obtained
as
\begin{equation}
g_{12}^{(2)}=g_{21}^{(2)}=g_{22}^{(2)}
\end{equation}
 and

\begin{align}
g_{11}^{(2)} & =\frac{n(1+D_{a})}{4}\text{Tr}\left\{ \left[\left(\mathbf{I}+\frac{D_{a}\tau_{p}}{1+D_{a}}\mathbf{\Lambda}\right)^{-1}\mathbf{\Lambda}^{-1}\right]^{2}\mathbf{\Lambda}^{-1}\right\} \\
 & +\frac{nD_{a}\tau_{p}}{4}\text{Tr}\left\{ \left[\left(\mathbf{I}+\frac{D_{a}\tau_{p}}{1+D_{a}}\mathbf{\Lambda}\right)^{-1}\mathbf{\Lambda}^{-1}\right]^{2}\right\} .
\end{align}

To evaluate the matrix inverse expressions analytically, we make use
of the Sherman--Morrison formula \citep{sherman1950adjustment}:
\begin{equation}
(\boldsymbol{A}+\boldsymbol{u}\boldsymbol{v}^{T})^{-1}=\boldsymbol{A}^{-1}-\frac{\boldsymbol{A}^{-1}\boldsymbol{u}\boldsymbol{v}^{T}\boldsymbol{A}^{-1}}{1+\boldsymbol{v}^{T}\boldsymbol{A}^{-1}\boldsymbol{u}}.
\end{equation}
 This identity holds for an invertible matrix $\boldsymbol{A}$ and
vectors $\boldsymbol{u}$, $\boldsymbol{v}$. Assuming $\boldsymbol{A}=a\mathbf{I}$
$(a\neq0)$, $\boldsymbol{u}=b\mathbf{1}$, and $\boldsymbol{v}=\mathbf{1}$,
the formula reduces to:
\begin{eqnarray}
\left(a\mathbf{I}+b\boldsymbol{1}\boldsymbol{1}^{T}\right)^{-1} & = & \frac{1}{a}\mathbf{I}-\frac{\frac{1}{a}\mathbf{I}b\mathbf{11}^{\text{T}}\frac{1}{a}\mathbf{I}}{1+\mathbf{1}^{\text{T}}\frac{1}{a}\mathbf{I}b\mathbf{1}}\nonumber \\
 & = & \frac{1}{a}\mathbf{I}-\frac{\frac{b}{a^{2}}\mathbf{11}^{\text{T}}}{1+\frac{b}{a}\mathbf{1^{T}1}^{\text{ }}}\nonumber \\
 & = & \frac{1}{a}\mathbf{I}-\frac{b}{a(a+bn)}\mathbf{11}^{\text{T}}.
\end{eqnarray}
 This relation provides the general form required to evaluate all
matrix inverse terms appearing in this section.

Although analytical expressions for the metric have been obtained,
the resulting geodesic equations are sufficiently complicated to require
numerical solution. The details of this standard numerical procedure
are omitted here for brevity.

\subsection{Pairwise Gaussian-core Interaction}

For interacting systems beyond quadratic potentials, analytical solutions
are generally unavailable. We reformulate the problem as a variational
optimization, which can be efficiently tackled using sampling-based
methods. Consider a system confined in a harmonic trap of strength
$\lambda_{1}(t)$, with the particles $i$ and $j$ coupled via a
pairwise interaction potential 
\begin{equation}
u_{ij}=\lambda_{2}(t)\exp(-\frac{r_{ij}^{2}}{2\sigma^{2}}),
\end{equation}
 where $r_{ij}\equiv|\boldsymbol{r}_{i}-\boldsymbol{r}_{j}|$ is the
interparticle distance. The total potential energy of the system is
therefore: 
\[
U_{o}=\frac{1}{2}\lambda_{1}\boldsymbol{r}_{i}^{2}+\frac{1}{2}\lambda_{2}(t)\sum_{i\neq j}\exp(-\frac{r_{ij}^{2}}{2\sigma^{2}}).
\]

We begin by retaining terms up to first order in $\tau_{p}$. In this
framework, the auxiliary potential $U_{a}^{(1)}$, which steers the
system along the distribution $\rho_{B}(\boldsymbol{r},\boldsymbol{\lambda}(t))\equiv\exp\{\frac{F(\boldsymbol{\lambda}(t))-U_{o}(\boldsymbol{r},\boldsymbol{\lambda}(t))}{1+D_{a}}\}$,
is determined by the equation: 
\begin{equation}
(1+D_{a})\nabla^{2}U_{a}-\nabla_{i}U_{a}\cdot\nabla_{i}U_{o}=\frac{dF}{dt}-\frac{\partial U_{o}}{\partial t}.
\end{equation}
 Because $U_{o}$ is sufficiently complex to preclude an analytical
solution, and the many-particle nature of the system leads to the
curse of dimensionality in traditional grid-based PDE solvers (e.g.,
finite difference \citep{smith1985numerical} or finite element methods
\citep{brenner2008mathematical}), we seek to transform the problem
into a variational formulation amenable to sampling methods. Substituting
$U_{a}^{(1)}=\dot{\lambda}_{1}f_{1}+\dot{\lambda}_{2}f_{2}$ yields:
\begin{equation}
\dot{\lambda}_{1}\left\{ (1+D_{a})\nabla^{2}f_{1}-\nabla_{i}f_{1}\cdot\nabla_{i}U_{o}+\frac{\partial U_{o}}{\partial\lambda_{1}}-\frac{\partial F}{\partial\lambda_{1}}\right\} +\dot{\lambda}_{2}\left\{ (1+D_{a})\nabla^{2}f_{2}-\nabla_{i}f_{2}\cdot\nabla_{i}U_{o}+\frac{\partial U_{o}}{\partial\lambda_{2}}-\frac{\partial F}{\partial\lambda_{2}}\right\} =0.\label{eq:U_a_first_order}
\end{equation}
 We introduce two functionals: 
\begin{equation}
G_{1}(f_{1})\equiv\int\text{d}\boldsymbol{r}\left[(1+D_{a})\nabla^{2}f_{1}-\nabla_{i}f_{1}\cdot\nabla_{i}U_{o}+\frac{\partial U_{o}}{\partial\lambda_{1}}-\frac{\partial F}{\partial\lambda_{1}}\right]^{2}\exp\left(-\frac{U_{o}}{1+D_{a}}\right)
\end{equation}
 and 
\begin{equation}
G_{2}(f_{2})\equiv\int\text{d}\boldsymbol{r}\left[(1+D_{a})\nabla^{2}f_{2}-\nabla_{i}f_{2}\cdot\nabla_{i}U_{o}+\frac{\partial U_{o}}{\partial\lambda_{2}}-\frac{\partial F}{\partial\lambda_{2}}\right]^{2}\exp\left(-\frac{U_{o}}{1+D_{a}}\right).
\end{equation}
 The stationary conditions
\begin{equation}
\delta G_{1}(f_{1})/\delta f_{1}=0
\end{equation}
 and 
\begin{equation}
\delta G_{2}(f_{2})/\delta f_{2}=0
\end{equation}
 correspond exactly to Eq. (\ref{eq:U_a_first_order}). This construction
transforms the original PDE into an optimization problem over function
space. A key challenge is that $F$ is unknown. We can transform $G_{1}(f_{1})$
and $G_{2}(f_{2})$ so that their variational equations remain equivalent
to the target equation without explicit knowledge of $F$. Consider
$G_{1}(f_{1})$ as an example: 
\begin{eqnarray*}
G_{1}(f_{1}) & = & \int\text{d}\boldsymbol{r}D_{1}^{2}(f_{1})\exp\left(-\frac{U_{o}}{1+D_{a}}\right)\\
 & = & \int\text{d}\boldsymbol{r}\left[(1+D_{a})\nabla^{2}f_{1}-\nabla_{i}f_{1}\cdot\nabla_{i}U_{o}\right]^{2}\exp\left(-\frac{U_{o}}{1+D_{a}}\right)\\
 &  & +2(1+D_{a})\int\text{d}\boldsymbol{r}\left(\frac{\partial U_{o}}{\partial\lambda_{1}}-\frac{\partial F}{\partial\lambda_{1}}\right)\nabla_{i}\cdot\left[\left(\nabla_{i}f_{1}\right)e^{-\frac{U_{o}}{1+D_{a}}}\right]\\
 &  & +\int\text{d}\boldsymbol{r}\left(\frac{\partial U_{o}}{\partial\lambda_{1}}-\frac{\partial F}{\partial\lambda_{1}}\right)^{2}\exp\left(-\frac{U_{o}}{1+D_{a}}\right).
\end{eqnarray*}
The third term is independent of $f_{1}$ and can be omitted without
affecting the variational result. Applying integration by parts to
the second term and assuming boundary terms vanish, we obtain
\begin{eqnarray*}
\int\text{d}\boldsymbol{r}\left(\frac{\partial U_{o}}{\partial\lambda_{1}}-\frac{\partial F}{\partial\lambda_{1}}\right)\nabla_{i}\cdot\left[\left(\nabla_{i}f_{1}\right)e^{-\frac{U_{o}}{1+D_{a}}}\right] & = & -\int\text{d}\boldsymbol{r}\nabla_{i}\left(\frac{\partial U_{o}}{\partial\lambda_{1}}-\frac{\partial F}{\partial\lambda_{1}}\right)\cdot\left(\nabla_{i}f_{1}\right)e^{-\frac{U_{o}}{1+D_{a}}}\\
 & = & -\int\text{d}\boldsymbol{r}\left(\nabla_{i}\frac{\partial U_{o}}{\partial\lambda_{1}}\right)\cdot\left(\nabla_{i}f_{1}\right)e^{-\frac{U_{o}}{1+D_{a}}},
\end{eqnarray*}
where the last line follows because $F$ is independent of spatial
coordinates. Consequently, we can define the modified functionals:
\begin{eqnarray}
G_{1}(f_{1}) & \equiv & \langle\left[(1+D_{a})\nabla^{2}f_{1}-\nabla_{i}f_{1}\cdot\nabla_{i}U_{o}\right]^{2}\nonumber \\
 &  & -2(1+D_{a})\nabla_{i}f_{1}\cdot\nabla_{i}\frac{\partial U_{o}}{\partial\lambda_{1}}\rangle,\label{eq:f_1_functional}
\end{eqnarray}

\begin{eqnarray}
G_{2}(f_{2}) & \equiv & \langle\left[(1+D_{a})\nabla^{2}f_{2}-\nabla_{i}f_{2}\cdot\nabla_{i}U_{o}\right]^{2}\nonumber \\
 &  & -2(1+D_{a})\nabla_{i}f_{2}\cdot\nabla_{i}\frac{\partial U_{o}}{\partial\lambda_{2}}\rangle.\label{eq:f_2_functional}
\end{eqnarray}
 The stationary conditions $\delta G_{1}(f_{1})/\delta f_{1}=0$ and
$\delta G_{2}(f_{2})/\delta f_{2}=0$ derived from Eqs. (\ref{eq:f_1_functional})
and (\ref{eq:f_2_functional}) are equivalent to Eq. (\ref{eq:U_a_first_order}).
We now adopt the following ansatzes to parameterize $f_{1}$ and $f_{2}$:
\begin{equation}
f_{1}(\boldsymbol{r},\boldsymbol{\lambda})=a(\boldsymbol{\lambda})\sum_{i\neq j}r_{ij}^{2}+b(\boldsymbol{\lambda})\boldsymbol{r}_{i}^{2}
\end{equation}
 and

\begin{equation}
f_{2}(\boldsymbol{r},\boldsymbol{\lambda})=c(\boldsymbol{\lambda})\sum_{i\neq j}r_{ij}^{2}+d(\boldsymbol{\lambda})\boldsymbol{r}_{i}^{2}.
\end{equation}
 We adopt a low-order polynomial ansatz, which captures pair correlations
while remaining computationally tractable. Performing the variational
procedure with respect to the coefficients leads to a system of linear
equations for $a$, $b$, $c$, $d$, which can then be solved straightforwardly.
The subsequent derivation of the auxiliary potential, the metric tensor,
and corresponding geodesic paths are described in the main text.

Next, we retain terms up to second order in $\tau_{p}$. The auxiliary
potential is now expressed as $U_{a}^{(2)}=u+\dot{\boldsymbol{\lambda}}\cdot\boldsymbol{\omega}$,
where $u$ and \textbf{$\boldsymbol{\omega}\equiv\left(\omega_{1},\omega_{2}\right)$}
satisfy
\begin{equation}
D_{a}\tau_{p}\frac{\partial}{\partial r_{j}^{\beta}}\left(\frac{\partial^{2}(U_{o}+u)}{\partial r_{i}^{\alpha}\partial r_{j}^{\beta}}\rho_{B}\right)=\frac{\partial(U_{o}+u)}{\partial r_{i}^{\alpha}}\rho_{B}+(1+D_{a})\frac{\partial\rho_{B}}{\partial r_{i}^{\alpha}}
\end{equation}
 and 
\begin{equation}
\frac{\partial\rho_{B}}{\partial\lambda_{\mu}}=\frac{\partial}{\partial r_{i}^{\alpha}}\left(\frac{\partial\omega_{\mu}}{\partial r_{i}^{\alpha}}\rho_{B}\right)-D_{a}\tau_{p}\frac{\partial^{2}}{\partial r_{i}^{\alpha}\partial r_{j}^{\beta}}\left(\frac{\partial^{2}\omega_{\mu}}{\partial r_{i}^{\alpha}\partial r_{j}^{\beta}}\rho_{B}\right).
\end{equation}
Simplifying yields 
\begin{equation}
D_{u}^{(2)}(u)\equiv\frac{\partial u}{\partial r_{i}^{\alpha}}-D_{a}\tau_{p}\frac{\partial}{\partial r_{j}^{\beta}}\frac{\partial^{2}(U_{o}+u)}{\partial r_{i}^{\alpha}\partial r_{j}^{\beta}}+\frac{D_{a}\tau_{p}}{1+D_{a}}\frac{\partial^{2}(U_{o}+u)}{\partial r_{i}^{\alpha}\partial r_{j}^{\beta}}\frac{\partial U_{o}}{\partial r_{j}^{\beta}}=0\label{eq:u_equation_gaussian}
\end{equation}
 and 
\begin{eqnarray}
D_{\mu}^{(2)}(\omega_{\mu}) & \equiv & (1+D_{a})\nabla^{2}\omega_{\mu}-\nabla_{i}\omega_{\mu}\cdot\nabla_{i}U_{o}+\frac{\partial U_{o}}{\partial\lambda_{\mu}}-\frac{\partial F}{\partial\lambda_{\mu}}\nonumber \\
 & - & (1+D_{a})D_{a}\tau_{p}\nabla^{4}\omega_{\mu}+2D_{a}\tau_{p}\nabla_{i}\nabla^{2}\omega_{\mu}\cdot\nabla_{i}U_{o}\nonumber \\
 & - & D_{a}\tau_{p}\frac{\partial^{2}\omega_{\mu}}{\partial r_{i}^{\alpha}\partial r_{j}^{\beta}}\left(\frac{1}{1+D_{a}}\frac{\partial U_{o}}{\partial r_{i}^{\alpha}}\frac{\partial U_{o}}{\partial r_{j}^{\beta}}-\frac{\partial^{2}U_{o}}{\partial r_{i}^{\alpha}\partial r_{j}^{\beta}}\right)=0.\label{eq:omega_mu_equation}
\end{eqnarray}
 We define two functionals: 
\begin{equation}
G_{u}^{(2)}(u)\equiv\langle[D_{u}^{(2)}(u)]^{2}\rangle\label{eq:functional_u}
\end{equation}
 and 
\begin{equation}
G_{\mu}^{(2)}(\omega_{\mu})\equiv\langle[D_{\mu}^{(2)}(\omega_{\mu})]^{2}\rangle,\label{eq:functional_omega}
\end{equation}
 whose stationary conditions 
\begin{equation}
\delta G_{u}^{(2)}(u)/\delta u=0
\end{equation}
 and 
\begin{equation}
\delta G_{\mu}^{(2)}(\omega_{\mu})/\delta\omega_{\mu}=0
\end{equation}
 are equivalent to Eqs. (\ref{eq:u_equation_gaussian}) and (\ref{eq:omega_mu_equation}),
respectively. Note that the term $\frac{\partial F}{\partial\lambda_{\mu}}$
in Eq. (\ref{eq:omega_mu_equation}) is unknown; we therefore use
the identity $\frac{\partial F}{\partial\lambda_{\mu}}=\langle\frac{\partial U_{o}}{\partial\lambda_{\mu}}\rangle$
which follows from 
\begin{eqnarray*}
\frac{\partial F}{\partial\lambda_{\mu}} & = & -(1+D_{a})\frac{\partial}{\partial\lambda_{\mu}}\log\int\text{d}\boldsymbol{r}\exp\left(-\frac{U_{o}}{1+D_{a}}\right)\\
 & = & \frac{\int\text{d}\boldsymbol{r}\frac{\partial U_{o}}{\partial\lambda_{\mu}}\exp\left(-\frac{U_{o}}{1+D_{a}}\right)}{\int\text{d}\boldsymbol{r}\exp\left(-\frac{U_{o}}{1+D_{a}}\right)}\\
 & = & \langle\frac{\partial U_{o}}{\partial\lambda_{\mu}}\rangle.
\end{eqnarray*}
We adopt the following parameterizations for $u$, $\omega_{1}$ and
$\omega_{2}$: 
\begin{equation}
u=a(\boldsymbol{\lambda})\sum_{i\neq j}\exp\left[-\frac{r_{ij}^{2}}{2\sigma^{2}}\right]+b(\boldsymbol{\lambda})\boldsymbol{r}_{i}^{2},
\end{equation}
\begin{equation}
\omega_{1}=c(\boldsymbol{\lambda})\sum_{i\neq j}r_{ij}^{2}+d(\boldsymbol{\lambda})\boldsymbol{r}_{i}^{2},
\end{equation}

\begin{equation}
\omega_{2}=e(\boldsymbol{\lambda})\sum_{i\neq j}r_{ij}^{2}+f(\boldsymbol{\lambda})\boldsymbol{r}_{i}^{2}.
\end{equation}
Applying the variational procedure with respect to the coefficients
$(a,b,c,d,e,f)$ leads to a system of linear equations that can be
solved directly. The subsequent derivation of the auxiliary potential,
the metric tensor and the corresponding geodesic paths is also provided
in the main text.

These examples demonstrate that our framework applies seamlessly to
systems ranging from exactly solvable models to strongly interacting
many-body systems, providing both analytical insight and practical
computational schemes.

\bibliographystyle{apsrev4-1}
\bibliography{SM_bibtex}